\documentclass[fleqn,usenatbib]{mnras}

\usepackage{newtxtext,newtxmath}
\usepackage{multirow}
\usepackage{comment}

\usepackage[T1]{fontenc}
\usepackage{ae,aecompl}

\usepackage{graphicx}	
\usepackage{amsmath}	
\usepackage{amssymb}	

\title[Fly-by encounters between two planetary systems II]{Fly-by encounters between two planetary systems II: Exploring the interactions of diverse planetary system architectures}

\author[Li, Mustill \& Davies]{
Daohai Li\thanks{E-mail: li.daohai@astro.lu.se, lidaohai@gmail.com (DL)},
Alexander J. Mustill,
and Melvyn B. Davies
\\
Lund Observatory, Department of Astronomy and Theoretical Physics, Lund University, Box 43, SE-221 00 Lund, Sweden
}

\date{Accepted XXX. Received YYY; in original form ZZZ}

\pubyear{2015}

\begin{document}
\label{firstpage}
\pagerange{\pageref{firstpage}--\pageref{lastpage}}
\maketitle

\begin{abstract}
Planetary systems formed in clusters may be subject to stellar encounter flybys. Here we create a diverse range of representative planetary systems with different orbital scales and planets' masses and examine encounters between them in a typical open cluster. We first explore the close-in multi-super earth systems $\lesssim0.1$ au. They are resistant to flybys in that only ones inside a few au can destabilise a planet or break the resonance between such planets. But these systems may capture giant planets onto wide orbits from the intruding star during distant flybys. If so, the original close-in small planets' orbits may be tilted together through Kozai--Lidov mechanism, forming a ``cold'' system that is significantly inclined against the equator of the central host. Moving to the intermediately-placed planets around solar-like stars, we find that the planets' mass gradient governs the systems' long-term evolution post-encounter: more massive planets have better chances to survive. Also, a system's angular momentum deficit, a quantity describing how eccentric/inclined the orbits are, measured immediately after the encounter, closely relates to the longevity of the systems -- whether or not and when the systems turn unstable in the ensuing evolution millions of years post-encounter. We compare the orbits of the surviving planets in the unstable systems through (1) the immediate consequence of the stellar fly or (2) internal interplanetary scattering long post-encounter and find that those for the former are systematically colder. Finally, we show that massive wide-orbit multi-planet systems like that of HR 8799 can be easily disrupted and encounters at a few hundreds of au suffice.
\end{abstract}



\defcitealias{Li2019}{Paper I}

\section{Introduction}

Planetary systems together with their host stars are subject to stellar flybys. Most of such encounters, especially if the planetary systems are in the field, are probably weak because of the fast relative velocity \citep{Veras2012}. However, if in a clustered environment, where most stars are born \citep{Lada2003}, stellar encounters can be of low relative velocity are thus of particular importance \citep[for example,][]{Li2015}.

We consider open clusters in our work. As the name suggests, open clusters gradually dissolve, due to, e.g., tidal truncation by the galactic potential. Large clusters can hold together for longer \citep{Lamers2006} while small ones may evaporate quickly even before the member stars can interact with each other \citep{Adams2001a}. The clusters accounted for here have several hundreds to a few hundreds of stars. Such clusters have lifetimes of several hundreds of millions of years \citep{Lamers2006}, allowing the stars to sufficiently gravitationally interact with each other. Such clusters are not too dense and the planet formation process is not affected much inside tens of au \citep{Adams2006,Winter2018}. For instance, the solar system may originate from a cluster of a few thousands of stars \citep[see][and references therein]{Adams2010}. Clusters of this sizes, while less massive than the larger ones, are more numerous and contribute a similar of stars to the Galaxy \citep{Lada2003}. In such clusters, an average member star is expected to experience an encounter inside of a few hundreds of au \citep{Adams2006}. At a relative velocity of only a few km/s, these encounters are highly gravitationally focused \citep{Binney2008}. As a result, the encounter distance $r_\mathrm{enc}$ has a flat probability distribution function -- close encounters are as likely as distant ones \citep{Malmberg2011}. For instance, $\sim$ 20\% of solar-mass stars encounter another star within 100 au in such an environment \citep{Malmberg2007}.

Exoplanets have been observed around various types of star and in various environments \citep[e.g., in the Hyades open cluster,][]{Livingston2018}. And it has been estimated that the occurrence rate in open cluster M 67 is similar to that in the field \citep{Brucalassi2017}. This seems to suggest that planetary formation process is not disrupted much within a few tens of au in open clusters \citep[for example,][]{Adams2006,Pfalzner2018a,Winter2018}.

Then we can now worry about the effect of the aforementioned close encounters on the already-formed planets. It turns out that though infrequent they may be highly detrimental. For example, those with $r_\mathrm{enc}$ comparable to the scale of the orbits of the planets can lead to strong disturbance, causing their immediate ionisation \citep{Laughlin1998} and/or gravitational capture by the intruding star \citep{Malmberg2011}. And additionally, a multi-planet system, if mildly perturbed by the encounters, may develop instability millions of years after the encounter \citep{Malmberg2011}.

In \citet[][hereafter \citetalias{Li2019}]{Li2019}, we simulated encounters between a solar system with the four giant planets and a single Sun (type 1 encounter) or those between two solar systems (type 2). Two phases of simulations were looked into. In the first, dubbed as the encounter phase, the short-term immediate effect of the encounter was examined and in the second, the post-encounter phase, the planetary systems were propagated up to 1 Gyr, thus long-term evolution revealed. In the encounter phase, planets could be ejected as free floaters or be captured by the intruding star; but whether or nor the intruder itself already possessed a planetary system did not affect the outcome -- so the interplanetary interactions were negligible in this phase. In the post-encounter evolution however, the interplanetary forcing drives systems to further evolve. A good fraction of systems, though stable in the former phase, now showed strong scattering, leading to ejection of planets. While the less massive ice giants were the most vulnerable, Jupiter was the most resistant. The captured planets intensified the scattering and they themselves were lost to a large degree. When a capture survived, it ended up on a retrograde orbit with a chance of 50\%.

Here, as a followup of \citetalias{Li2019}, we explore a more diverse range of planetary systems, varying mainly the planets' stellar-centric distances and their masses. The former is directly related to how much perturbation the planets acquire during the encounter and the latter has to do with the later post-encounter interplanetary interactions.

The paper is organised as follows. In Section \ref{sec-planet-config}, we detail how the representative planetary configurations are chosen and numerically assembled. Then in Section \ref{sec-enc-config}, we describe our simulation setup, i.e., how we pick systems from Section \ref{sec-planet-config} and let them encounter. Sections \ref{sec-close-in}, \ref{sec-solar} and \ref{sec-hr8799} are devoted to presenting the results and discussing the implications for close-in, intermediately-placed and massive wide-orbit planets. Finally, Section \ref{sec-con} concludes the paper, listing the major findings.

\section{Planetary system configurations} \label{sec-planet-config}
Exoplanets are ubiquitously observed and their occurrence rates around different types of stars have been presented \citep{Winn2015}. For example, it was estimated that a few tens of per cent of solar-like stars have close-in small planets \citep{Howard2010,Zhu2018} and at a somehow lower rate, gas giants at a few au \citep{Cumming2008,Bryan2016,Wittenmyer2016}. Here we study three types of planetary systems, close-in planets within a few 0.1 to 1 au, those intermediately-placed at a few to a few tens of au and those massive and on even wider orbits.

\subsection{Close-in planets}
We first describe how the close-in planets are created. These planets have short orbital periods.

M-dwarfs are the most common stars. Each such star, on average, has $\sim$2 close-in small planets with orbital periods from a few to a few hundreds of days \citep{Dressing2015,Gaidos2016,Tuomi2019}. These planets with radii a few times of that of the Earth will be referred to as super earths in the remaining of the paper. Many of these M-dwarf systems have been detected by the {\it Kepler} mission. When plotting the period ratio of adjacent planets in {\it Kepler}-multis, an imbalance was found near the first order mean motion resonances (MMRs), forming a peak just wide of the MMR and a dip at the opposite side \citep{Fabrycky2014,Winn2015}. Different models have been proposed to explain this feature, for example, a combination of tidal damping and resonant forcing \citep{Lithwick2012}.

Here we do not use observed systems but create a representative M-dwarf two-planet system. The central star is of 0.5 solar mass and a super earth is of 3 earth masses. We first construct 1-planet systems to test the chance for ionisation. Two such systems are generated, one with a super earth at 0.2 au (corresponding to an orbital period of $\sim 50$ days) and the other with one at 0.33 au ($\sim 100$ days). These two are called ``MS1'' and ``MS2'', respectively. We then construct two-planet systems, one set in and the other close to the 2:1 MMR. In the first, the two planets are migrated into the MMR such that the inner planet is at $\sim$ 0.2 au and the outer at $\sim$ 0.33 au using the prescription of \citet{Lee2002}; this referred to as ``MDR'' (M-dwarf double resonant). In the second, the outer planet is randomly placed such that the resulting period ratio between it and the inner (fixed at 0.2 au) is within $(1.8,2.2)$; this is the ``MD'' system. Clearly, some of the MD systems are actually in the MMR.

Then, we would like to investigate planetary systems that more closely resemble the observed exoplanets. Solar type stars with close-in planets on average have three close-in super earths with orbital periods less than a few hundreds of days \citep{Zhu2018}. As will be seen from the M-dwarf systems, such compact systems are fairly safe from encounter flybys, and planets on wider orbits may make the system more vulnerable. In general, around solar-type stars, the occurrence rate for giant planets at a few au is $\sim$10\% \citep{Cumming2008,Wittenmyer2016}. Nonetheless, it was recently realised that close-in super earths and outer giants might be positively correlated \citep{Zhu2018a,Bryan2019} and as a result, systems containing the former is more likely to harbour the latter \citep[c.f.,][]{Bryan2016}.

Here we use the Kepler-48 system as a prototype. There, a 0.9 solar-mass K star hosts 3 close-in transiting planets (b, c and d, from inside out) with orbital periods $<$ 50 days \citep{Steffen2013,Marcy2014} and a radial velocity (RV) giant planet (e) on a 1000-day trajectory \citep{Marcy2014}. The inner two planets are close to the 2:1 MMR, generating appreciable transiting time variations \citep{Steffen2013}. The masses of the planets have been measured by RV and the planets' orbital periods and thus semimajor axes are also well constrained. But this is not the case for the eccentricity $e$ or inclination $i$. Here we consider a cold system in that $e$ and $i$ is randomly drawn from a Rayleigh distribution with dispersion 0.01 (rad) and the phase angles are random for all four planets.

\subsection{Intermediately-placed planets}
Then we move to planets on wider, intermediately-placed orbits, assumed to be revolving around Solar-like stars.

In the original solar system, the giant planet masses roughly decline with heliocentric distances. As \citet{Hao2013} and \citetalias{Li2019} showed, Jupiter, the most massive and also the one with the largest specific binding energy to the Sun, was the most stable during both the encounter and the post-encounter phases. In the first phase, the star-planet forcing dominates and the planets can be treated as test particles \citepalias{Li2019}, meaning that the durability of Jupiter is a result of its small heliocentric distance. However, in the second phase, the encounter is long finished and any dynamical phenomenon must be a direct consequence of interplanetary interactions. It is then not obvious whether it is the small semimajor axis or the large mass that mainly prescribes Jupiter's endurance. So here to study the effect of the radial mass gradient of the planets, we design a ``NUSJ'' system where the heliocentric distances of the solar system giant planets are simply reversed but their masses unchanged \citep[so Neptune is now at 5.2 au, Uranus at 9.6 au, Saturn at 19.1 au and Jupiter at 30.1 au; cf.][]{Raymond2010}. We have checked that these systems are stable over 100 Myr in isolation.

Additionally, we introduce a system with a flat mass slope, i.e., all the planets are equally-massed. Here a system with three Jupiter-massed planets, as was often used in planet scattering experiments \citep{Chambers1996,Marzari2002,Chatterjee2008}, is generated. In such a setup, the two neighbouring planets' separation is often measured in their mutual Hill radius, defined as \citep{Chambers1996}
\begin{equation}
R_{j,j+1}=\sqrt[3]{m_{j}+m_{j+1}\over 3 m_*}{a_{j}+a_{j+1}\over 2}
\end{equation}
where $m_{j}$ and $a_{j}$ are the mass and semimajor axis of the $j$th planet (or the $j+1$th if the subscripts are j+1); $m_*$ is the mass of the central star. And the two planets are often placed such that their mutual distance fulfils
\begin{equation}
\label{eq-k}
d_{j,j+1}=a_{j+1}-a_{j}=K R_{j,j+1}
\end{equation}
in which $K$ is a constant.

For a two-planet system, if both planets are on initially circular and coplanar orbits and if $K\gtrsim2\sqrt{3}$, close encounter is impossible \citep{Gladman1993}. However, for a system with three or more planets, no analytical criterion is available. Various works have derived empirical instability times for different $K$s \citep[e.g.][]{Chambers1996}. In practice, a planetary system is stable for Gyr if $K\gtrsim 10$ \citep{Morrison2016,Smith2009} and this is often the case for {\it Kepler} systems \citep{Fabrycky2014}.

Here, aiming to examine the effect of encounters on a system during both the encounter and the post-encounter phases, these systems should be stable on their own. Thus we choose $K=10$. In this configuration, the innermost planet is at 5.2 au, the current position of Jupiter; as such, the outermost is at 33 au. One thing to note is that our planets all have small initial eccentricity and inclination $\sim0.01$ (rad) following a Rayleigh distribution, which may play a role in determining the system stability \citep{Zhou2007,Pu2015}. Hence, we first run test simulations to make sure the so-configured systems are stable for 100 Myr on its own. These are referred to as ``3J10'' systems. Additionally, we introduce another set of systems where $K=15$ (``3J15''). Here we fix the outermost planet to have an orbit of the same size of 33 au target as that in the 3J10 systems. As such, the innermost is at about 1.5 au.

\subsection{Wide-orbit massive planets}\label{sec-sys-hr8799}
Finally we move to massive wide-orbit planets.

Wide-orbit massive planets (beyond tens of au) are probably rare \citep[e.g.,][]{Beuzit2019} with an estimated occurrence rate, through direct imaging, of $\sim 1\%$ \citep{Bowler2016,Galicher2016}. Among the tens of these planetary mass objects detected so far \citep[NASA Exoplanet Archive \url{https://exoplanetarchive.ipac.caltech.edu/} as retrieved on 2 Sept 2019;][]{Akeson2013}, HR 8799 \citep{Marois2008,Marois2010} hosts an intriguing system of four massive wide-orbit planets \citep[but multiple gas giant systems at a few au around massive stars may be not rare; see][and references therein]{Trifonov2019}.

HR 8799 is an A5V star of about 1.5 solar masses. Four planets (b, c, d and e, from the outermost to the innermost), each above 5 Jovian masses, orbit the star on paths $\sim10-80$ au from the host. Such closely-packed massive planets fail even to fulfil the two body Hill stability criterion \citep[so $K>2\sqrt{3}$ is not satisfied,][]{Fabrycky2010}. Consequently, the long-term stability of the system is not guaranteed and usually lock into successive MMRs is invoked \citep[e.g.,][but see \citealt{Gotberg2016} for an opposite view]{Fabrycky2010}. For our purpose, i.e., to study how encounter flybys affect the system's stability, we must first obtain systems that are stable in isolation in the first place. This is done through migrating the planets into consecutive MMRs; see details below. To enhance the chance of success, we have assigned each planet a mass of 5 Jupiter masses, close to the lower end of the estimate \citep{Marois2010,Gozdziewski2014}. In this sense, we are not replicating the exact HR 8799 system but are using it as a template to create systems of multiple wide-orbit massive planets.

It is not clear which MMRs are actually involved in the HR 8799 system \citep{Fabrycky2010,Gozdziewski2014}. Hence, instead of requiring a four-body resonance, we simply migrate the planets so that all adjacent pairs have period ratios close to two. Also, since the orbits of the actual planets are not well-constrained \citep{Marois2010} and we are not reproducing precisely the observed system, we allow the orbital elements of our planets to vary within reasonable ranges. The systems are assembled as follows. First we randomly draw, from 12 au to 18 au, a target semimajor axis for the innermost planet, e. Then the four planets' initial positions and their migration and damping timescales are then assigned following \citet{Gozdziewski2014}. We then let the system evolve for 10 Myr under migration and damping. If planet e reaches the target position, we stop the simulation and check if all planet pairs are close to 2:1 MMRs. If yes, this system is run for a further 100 Myr, now without migration or damping, to test its long-term stability; only stable systems are kept. Otherwise, if the innermost planet does not arrive at the target location within 10 Myr, the simulation is abandoned. We note the fact that planet e is between 12 and 18 au implies that the outermost one, b, can be anywhere between $\sim$ 50 and 80 au. Finally, due to the large planetary masses, Jacobi coordinates are used in this set of simulations.

In this way, a total of 100 HR 8799 analogue systems that are stable for 100 Myr have been created. Then for each of these, we take 100 snapshots from the first 1 Myr of the 100 Myr integration with no migration/damping, leading to a total $100\times100=10^4$ of realisations.

\begin{figure}
\includegraphics[width=\columnwidth]{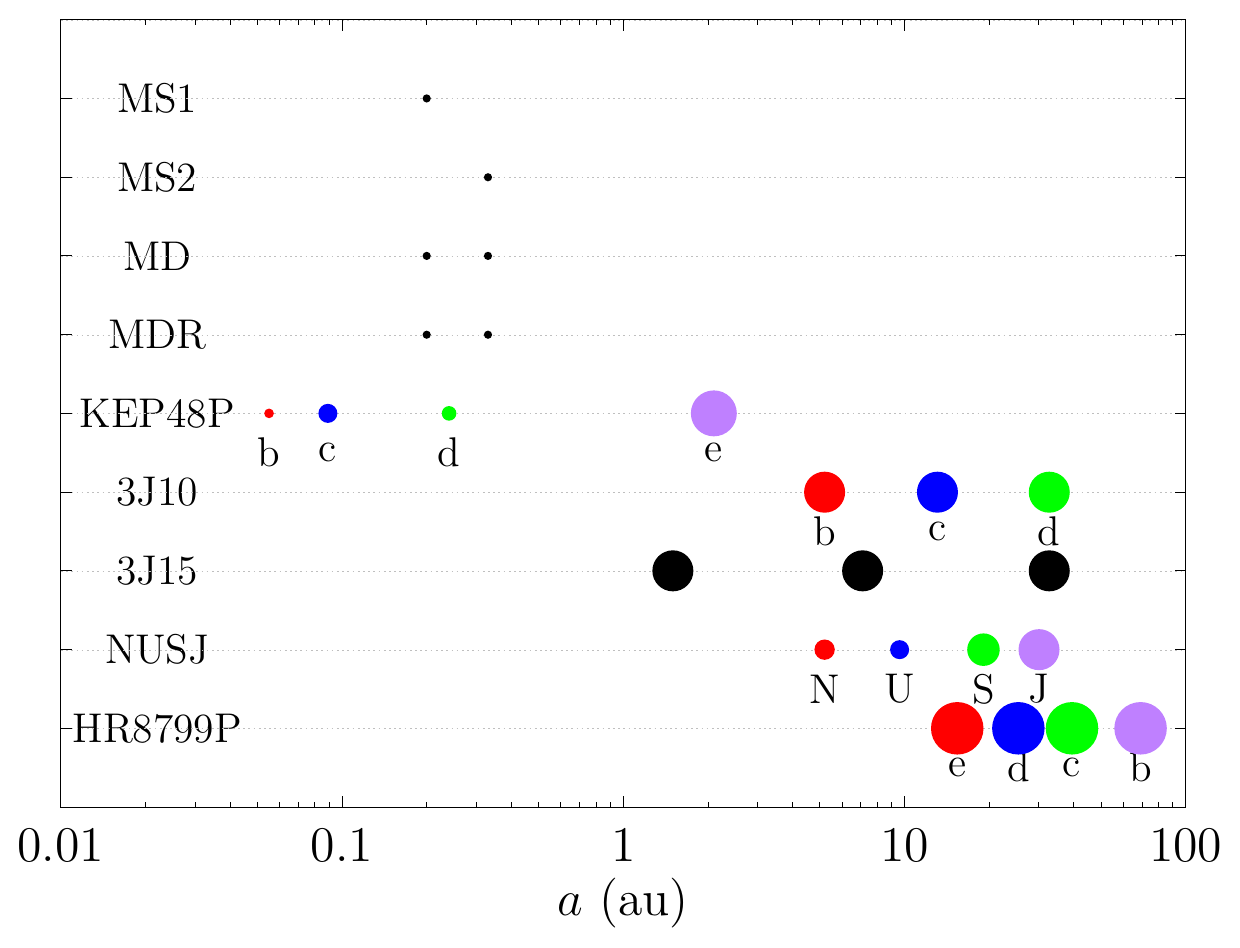}
\caption{Planetary systems that are tested in this work. Each point denotes a planet in a system, as labelled on the left of the row. The horizontal axis is the stellar-centric distance of the planet and the point size proportional to its logarithmic mass. All planets have small initial eccentricities (except HR 8799 e which has a value of $\sim 0.1$) and are largely coplanar. The HR 8799 system is scaled such that b is at 69 au. The planets' colours and labels correspond to those used later in Figures \ref{fig-cap-orb-Mdwarf}, \ref{fig-stab-3j10}, \ref{fig-stab-kep48-sun}, \ref{fig-ae-KP48-HR8799}, \ref{fig-cap-orb-KP48}, \ref{fig-stab-NUSJ} and \ref{fig-stab-hr8799}. See also Table \ref{tab-sys}.}
\label{fig-systems}
\end{figure}

All these systems described above are listed in Table \ref{tab-sys} and illustrated in Figure \ref{fig-systems} for ease of reference and they will be confronted with encounter flybys.

\begin{table}
\centering
\caption{Planetary systems that are tested in this work. In the three columns, we have the label of the system, the mass of the central star and the planetary system configuration; see text for details. Here $m_\odot$ is a Solar mass, $m_\oplus$ an Earth mass and $m_\mathrm{J}$ a Jovian mass; SE stands for a super earth (of a few to a few tens of earth masses). See also Figure \ref{fig-systems}.}
\label{tab-sys}
\begin{tabular}{c c c}
\hline
 label & stellar mass & planets \\
\hline
MS1 & 0.5 $m_\odot$ &1 SE, 3 $m_\oplus$, $\sim$0.2 au \\
MS2  & 0.5 $m_\odot$ &1 SE, 3 $m_\oplus$, $\sim$0.3 au \\
MD  & 0.5 $m_\odot$ & 2 SEs, each 3 $m_\oplus$, $\sim$0.2/0.3 au, near MMR\\
MDR  & 0.5 $m_\odot$ & 2 SEs, each 3 $m_\oplus$, $\sim$0.2/0.3 au, in MMR\\
KEP48P&0.9 $m_\odot$&3 SEs $\in(4,15)m_\oplus$, $\lesssim0.2$ au; 1 giant, 2$m_\mathrm{J}$, 2 au\\
3J10  & 1 $m_\odot$ & 3 Jupiters, $K=10$, $\in(5.2,33)$ au\\
3J15 & 1 $m_\odot$ & 3 Jupiters, $K=15$, $\in(1.5,33)$ au\\
NUSJ  & 1 $m_\odot$ & solar system giants in reverse order\\
HR8799P & 1.5 $m_\odot$ & 4 planets $\in(10,80)$ au, each 5 $m_\mathrm{J}$\\
\hline
\end{tabular}
\end{table}

\section{Encounter setup} \label{sec-enc-config}

The encounters between these planetary systems in a cluster can be studied either in a Monte Carlo approach in the sense that the encounters are directly generated according to the expected distribution \citep[e.g.,][]{Laughlin1998,Malmberg2011,Hao2013,Li2019} or in a self-consistent cluster simulation where thus the frequency and parameters of the flybys flybys are more accurately modelled \citep[see][for instance]{Spurzem2009,Cai2017,VanElteren2019,Fujii2019}. Here we follow \citetalias{Li2019} and adopt the first method. And similarly here we are only interested the encounters that can significantly change the planetary systems, potentially causing planet loss. This may occur in two phases, 1) immediate ejection of a planet during the encounter phase and 2) delayed instability during the post-encounter phase.

In the first case 1), direct ejection is only possible when the encounter distance is no larger than three times the planetary semimajor axis \citep{Pfalzner2005}. This critical distance only weakly depends on the mass of the intruder \citep{Bhandare2016,Jilkova2016}. For solar system planets, this implies that only encounters closer than $\sim90$ au may achieve this \citepalias{Li2019}.

In the second case 2) where the planetary system remains bound during the encounter and only becomes unstable later due to direct interplanetary interactions -- induced by and also long after the encounter \citep{Malmberg2011,Hao2013}. In \citetalias{Li2019}, we established that for the solar system, to exert this this type of delayed instability is not much easier than to immediately eject as described above and encounters $\lesssim$ 100 au are needed.

An additional complexity is that both encountering stars may have their own planetary systems. Now, one star (A) may, during the encounter, capture a planet from the encountering star (B). While this captured planet does not affect these indigenous to A during the encounter, in the later post-encounter evolution, it interacts with originals, causing a great reduction of system multiplicity \citepalias{Li2019}. Because it is a fraction of the planets ejected from the B that are captured by A, an encounter distance the same as that discussed in 1) is required.

\subsection{Encountering orbits}
The evidence gathered above suggests that only close encounters at distances no more than a few times the semimajor axes of the planets are of interest to this work. Now we describe how in simulations we choose the encounter distance for each of the planetary systems in Table \ref{tab-sys} and Figure \ref{fig-systems}.

All the four M-dwarf systems considered in this work are close-in and the planets are at $\sim$ 0.2/0.33 au. This then implies that only encounters inside $\sim$ 1 au may be relevant. Nonetheless, we note that MDR systems are in the 2:1 MMR and MD ones are also near/in the MMR. These configurations allow us to assess the survivability of MMRs under stellar flybys. This has been tested for possible planetary configurations of the early solar and it seems that to break a MMR chain is not much easier than to directly eject a planet \citep{Li2015}. Here to be conservative, we examine encounters with encounter distances up to $r_\mathrm{enc}<2$ au for these M-dwarf systems, twice as far as needed for immediate ejection. According to \citetalias{Li2019}, these encounters between a planetary system and a single star are type 1 encounters.

As said before, when captured into a planetary system, a planet may effect strong perturbation on the originals. Here we examine a case where the MD system encounters that of HR 8799. Two considerations are behind this choice: 1) the planets in the HR 8799 system are massive, hence potentially more capable of exerting disturbance and 2) the HR 8799 system is wide so capture may occur at large encounter distances (see below) and this minimises the effect of the stellar encounter itself. The scales of the two systems are different by two orders of magnitude. As a consequence, while to immediately eject a planet from the MD system, encounters closer than a few au is needed, to implant an additional planet into the MD system from the HR 8799 system, encounters at hundreds of au may be sufficient. Therefore, an upper limit of $r_\mathrm{enc}<300$ au is used here. These are the type 2 encounters \citepalias{Li2019}.

The outermost planet, e, of the Kepler-48 system is at $\sim 2$ au, so here we have $r_\mathrm{enc}<6$ au. The Kepler-48 systems are also tested against the HR 8799 systems in order to examine the effect of captures and then, a limit of 300 au is adopted, the same as that for the MD systems.

Next, we turn to planetary systems on intermediately placed orbits. Our 3J10 and 3J15 systems both have the outermost planet at $\sim 33$ au. Thus for these two, the maximum encounter distance is set to 100 au. As will be shown later, the NUSJ systems may be disrupted due to more distant encounters, so we allow $r_\mathrm{enc}$ to be 150 au for those systems.

Finally, we come to the HR 8799 systems. Because of the way of its construction, the semimajor axis of HR 8799 b, the outermost planet, has a large spreading in its semimajor axis, ranging from $\sim$ 50 to 80 au. Here, when confronting it with the Sun, we let $r_\mathrm{enc}<500$ au, meaning that the ratio of the maximum encounter distance to the semimajor axis of the outermost planet ranges roughly from 6 to 10 for different runs.
\begin{table}
\centering
\caption{Configurations of our encounters. The first column shows the target planetary system and the second the encountering star or planetary system. The third column is the maximum encounter distance and the fourth shows the length of the post-encounter simulation. The bottom two rows list the encounters between two planetary systems that are referred to as type 2 encounters whereas those above describe encounters between a planetary system and a single star, called type 1 encounters.}
\label{tab-enc}
\begin{tabular}{ c c c c }
\hline
target & flyer-by & $r_\mathrm{enc,max}$ (au) & length (Myr) \\
\hline
MS1 & Sun &  $2$ & -\\
MS2  & Sun &  $2$ & -\\
MD  & Sun &  $2$ &  1\\
MDR  & Sun &  $2$ &  1\\
KEP48P  & Sun &  $10$ &  1\\
3J10  & Sun &  $100$ &  100\\
3J15  & Sun &  $100$ &  100\\
NUSJ  & Sun &  $150$ &  100\\
HR8799P  & Sun &  $500$ &  100\\
\hline
MD  & HR8799P &  $300$ & 1\\
KEP48P  & HR8799P &  $300$ & 1\\
\hline
\end{tabular}
\end{table}

Our encounters have parameters typical for open clusters in the solar neighbourhood. Such clusters typically have hundreds of stars within a few parsecs aged hundreds of millions of years \citep{Kharchenko2013}. Here we adopt a single-valued velocity at infinity of $v_\infty=1$ km/s. In \citetalias{Li2019}, we showed that using a more realistic Maxwellian distribution \citep{Laughlin1998} did not change the result. This is because as long as $v_\infty$ is small, the encountering velocity at the closest approach is dominated by that converted from potential energy and this is the same for any $v_\infty$ at a given encounter distance.

\subsection{Integration times}

In all our type 1 encounters where a planetary system is visited by a single star, we are using the Sun as the flying-by star. On the other hand, because the purpose of these type 2 encounter is to study the influence of captured planets, only the MD/Kepler-48 systems that manage to capture a planet from the HR8799 system are propagated in the post-encounter phase. In addition, if the MD/Kepler-48 systems are left with only one planet after the encounter, that planet will follow the Koxai-Lidov cycles forced by the capture \citep{Kozai1962,Lidov1962,Naoz2013}. So another criterion for long-term integration is that the system should retain at least two original planets during the encounter.

As discussed above, an otherwise stable planetary system, if moderately disturbed by an encounter (so no immediate ejection), may turn unstable millions of years afterwards. In order to study this type of instability, all our simulations consist of two phases, a brief encounter phase and an extended post-encounter phase. While the length of the former is always $10^4$ yr, that of the latter is different for different planetary systems. Aiming to strike a balance between the length of a single run and the number of cases and the computational resources available, we have chosen, for the intermediately placed NUSJ, 3J10, 3J15 and the HR 8799 systems, an integration length of 100 Myr. For the close-in M-dwarf and the Kepler-48 systems, due to the short orbital periods (with closest planets on 50- and 5-day orbits, respectively), a simulation time of 1 Myr is used. \footnote{We note here that the $10^4$ yr time span for the encounter phase for especially the Kepler-48 system could be too long in the sense that it can be actually much longer than the secular timescales; the discussions are deferred to later sections.}

All our encounter phase simulations are composed of $10^4$ realisations for each encounter configuration. And the post-encounter simulations, being computationally more expensive, contain only $\sim$1000 integrations, randomly picked from the $10^4$. Or in the case of type 2 encounters, the number of post-encounter phase runs is the same as that of planet-capturing systems.

All encounters are listed in Table \ref{tab-enc} and they are all performed using the Bulirsch-Stoer integrator of the {\small MERCURY} $N$-body package \citep{Chambers1999} with an error tolerance of $10^{-12}$.

\section{Close-in planets} \label{sec-close-in}
We start by presenting our results with the close-in planets.
\subsection{MS1/2, MD and MDR: flyby by a single Sun}
\begin{figure}
\includegraphics[width=\columnwidth]{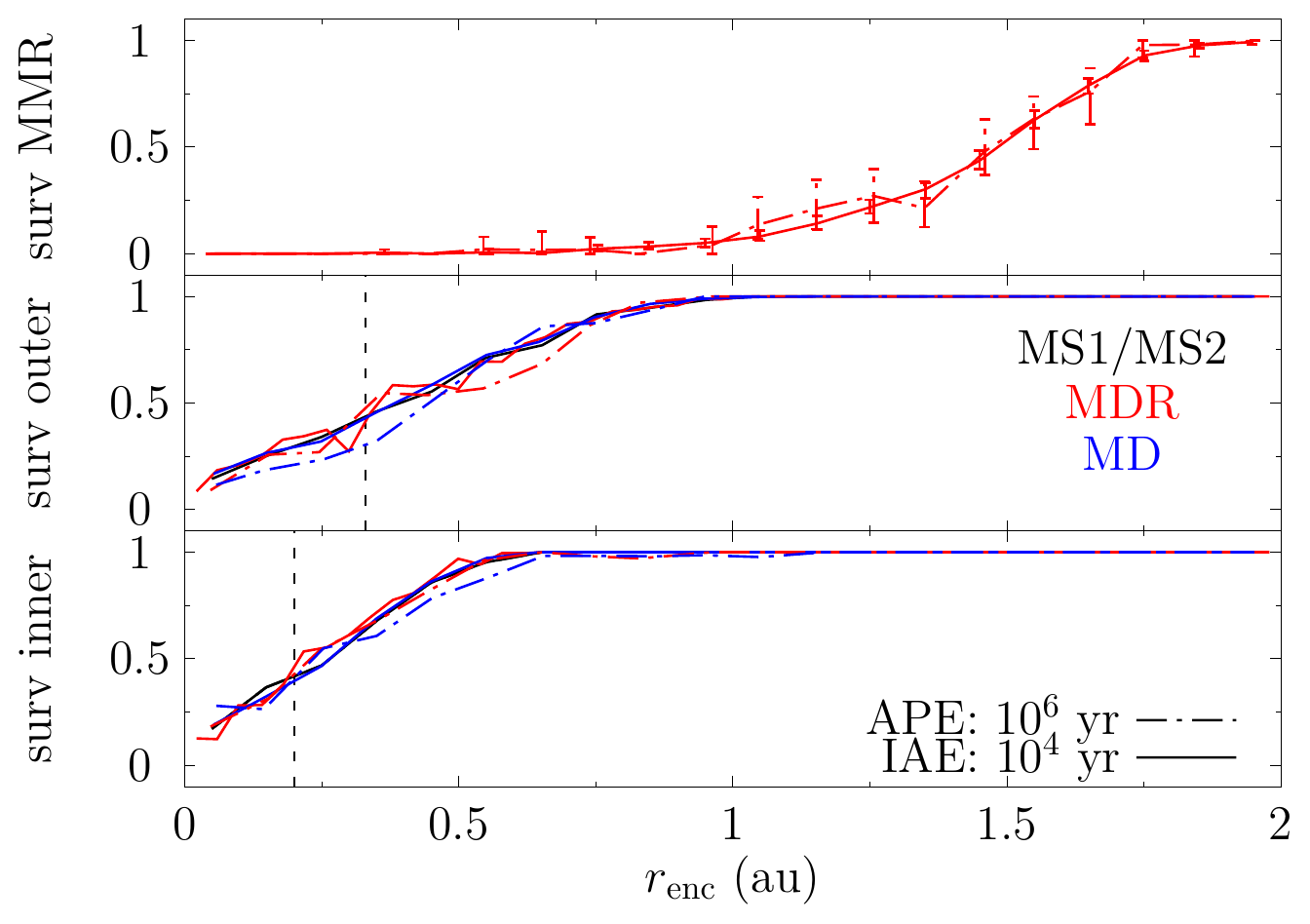}
\caption{Stability of the two planets and of the resonance as a function of the encounter distance at the two phases in the M-dwarf planetary systems encountering a single Sun. Bottom: survival fraction of the inner planet. Black, red and blue lines show MS1, MDR and MR simulations; solid and dash-dotted lines represent that immediately after the encounter (IAE) and after the post-encounter evolution (APE). Middle: the same as the bottom but for the outer planet. The vertical dashed lines mark the initial location of the planets. Top: fraction of MDR systems with the 2:1 MMR surviving, line styles the same as the bottom. The error bar is from a bootstrap resampling at 95\% confidence. It appears that the survivability at IAE can be higher than at APE; this results from that fact that for the former $10^4$ encounter phase runs are used whereas for the latter, only randomly selected $10^3$ cases that are propagated in the post-encounter phase contribute.}
\label{fig-stab-abc}
\end{figure}

We show in Figure \ref{fig-stab-abc} the survival fraction of the inner (bottom panel) and the outer (middle panel) planet as a function of encounter distance. The black, red and blue lines represent the MS1/2, MDR and MD systems, respectively. And the solid lines mean immediately after the encounter (IAE) at $10^4$ yr and the dashed vertical lines mark the planets' initial locations.

Reading from the bottom panel, for immediate ejection during the encounter of the inner planet at 0.2 au, an encounter as close as 0.6 au, or three times the planetary semimajor axis is needed \citep{Bhandare2016,Jilkova2016}. And, the three cases (MS1, MDR and MD) are indistinguishable, meaning that the interplanetary interactions are negligible during the brief encounter phase \citep[][and \citetalias{Li2019}]{Pfalzner2005a}. Thus as a consequence, the residence in or proximity to MMR does not play a role either. That for the outer planet is similar except that now the maximum encounter distance allowing for ejection moves to $\sim 1$ au.

The dash-dotted lines denote the stability after the post-encounter phase (APE) simulation. During this later evolution, few planets, shown as the difference between the solid and dash-dotted lines, are disrupted. Moreover, the inherent reason for such instability is uncertain -- all the 46/2000 of the MD and MDR systems losing a planet in this phase have become orbital crossing early on during the encounter. So in this sense, the only way to destabilise such a system is to do that during the encounter either by immediate ejection or by making the planets' orbits cross.

Then, in the top panel, we show the survival fraction of the 2:1 MMR for the MDR systems where the two planets have been initially migrated into MMR before the encounter. An algorithm adapted from \citet{Li2016}, able to find the resonating episodes is the critical angle if not constantly librating, is used for resonance detection. It turns out that no resonance can survive should the encounter be closer than 1 au. This threshold coincides with the maximum $r_\mathrm{enc}$ at which an encounter can eject the outer planet. Also, around 2 au, the MMR cannot be damaged anymore; this is about six times the semimajor axis of the outer planet. The fact that the solid and the dash-dotted lines agree within the error bar means that MMRs cannot be broken during the post-encounter evolution in such systems.

\begin{figure}
\includegraphics[width=\columnwidth]{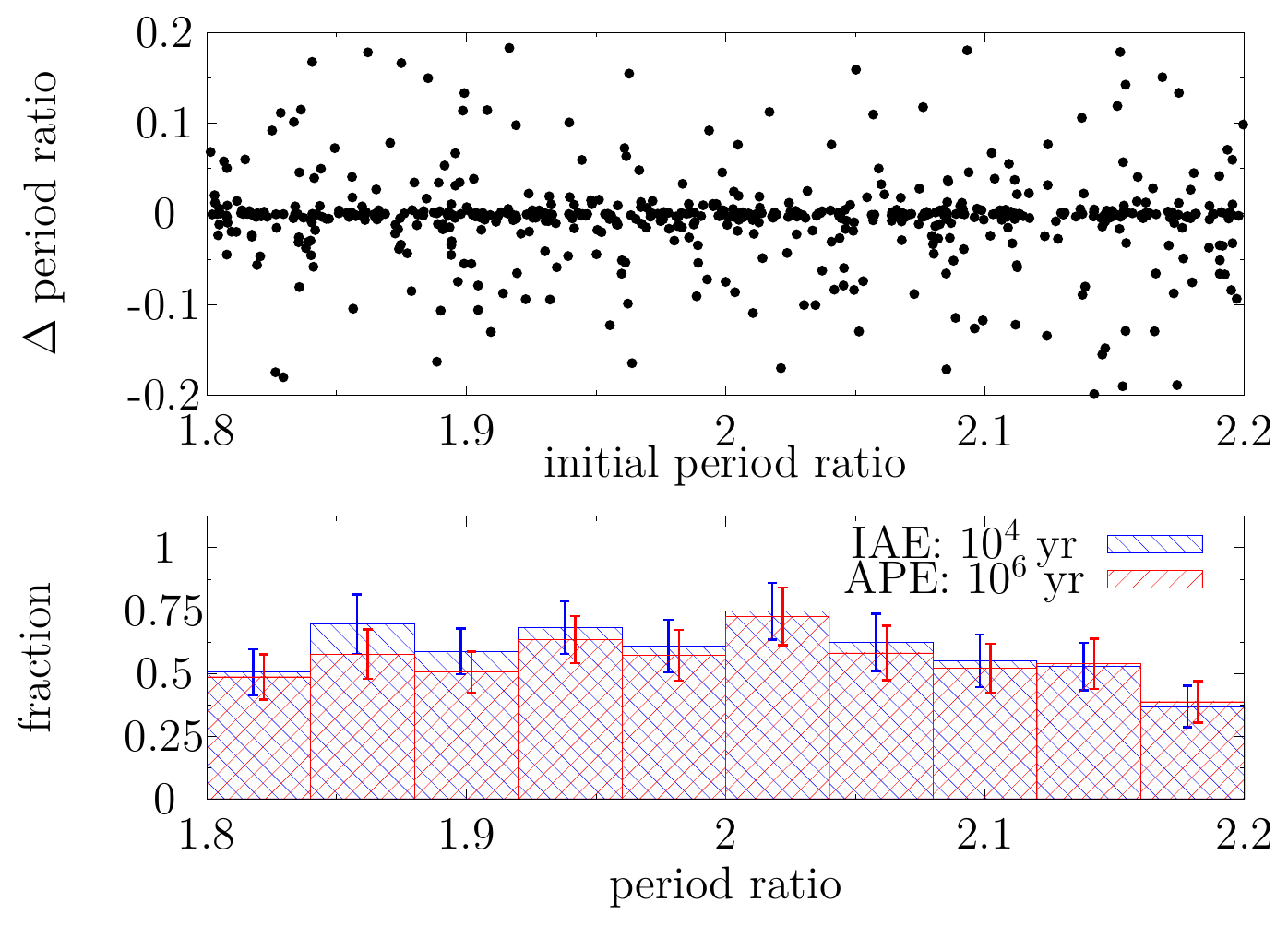}
\caption{Distribution of the period ratio between the two planets in the MD simulations. Bottom: the blue histogram shows the ratio immediately after the encounter (IAE) and the red that after the post-encounter evolution (APE), both normalised against the initial ratio before the encounter; the error bars are horizontally shifted for better visibility. Top: the change in the ratio at APE as a function of the initial value.}
\label{fig-resonance}
\end{figure}

Finally, as discussed earlier, the observed distribution of the ratio of orbital periods between adjacent exoplanets tend to be just wide of first order MMRs together with a deficit just inside of the exact resonance \citep{Fabrycky2014}. Here we wonder if the flybys preferentially push the period ratio in a specific direction. For this purpose, the MD simulations where the initial ratio before the encounter is evenly distributed around 2 are used. The bottom panel of Figure \ref{fig-resonance} shows the distribution of this ratio immediately after encounter (IAE) and that after the long-term evolution (APE), both normalised by the that before the encounter. It seems that at IAE there is a slight preference for period ratios $<2$. A random subsampling shows that at IAE, the relative frequency below and above the period ratio of 2 is $0.62\pm0.04$ and $0.56\pm0.04$; the two numbers at APE are $0.56\pm0.04$ and $0.55\pm0.04$. So while the there is a weak preference for smaller period ratio (opposite to the observations) at $1.5\sigma$ level, the post-encounter evolution smears that out. We suggest that stellar flybys are probably not a contributor to the observed imbalance.

\subsection{MD: flyby by HR 8799 system: effect of captured planets}\label{sec-mdwarf-hr8799}
Having established that the M-dwarf close-in planets are resistant against the flyby of a planetless Sun during both the encounter and the post-encounter phases in that only extremely close encounter ($\sim$ au) can cause destruction, we now proceed to examine the case where an additional massive planet is captured by the M-dwarf. This involves encounters with the HR 8799 systems, the so-called type 2 encounters.

In our $10^4$ encounter phase simulations, 381 planets are stolen by the M-dwarf from the HR 8799 system. Two or more planets can be captured simultaneously --  93 captures are hosted by 43 M-dwarf systems, each containing at least two such planets. The remaining 288 are lone captures, as the new host only grab one planet from the HR 8799 system.

\begin{figure}
\includegraphics[width=\columnwidth]{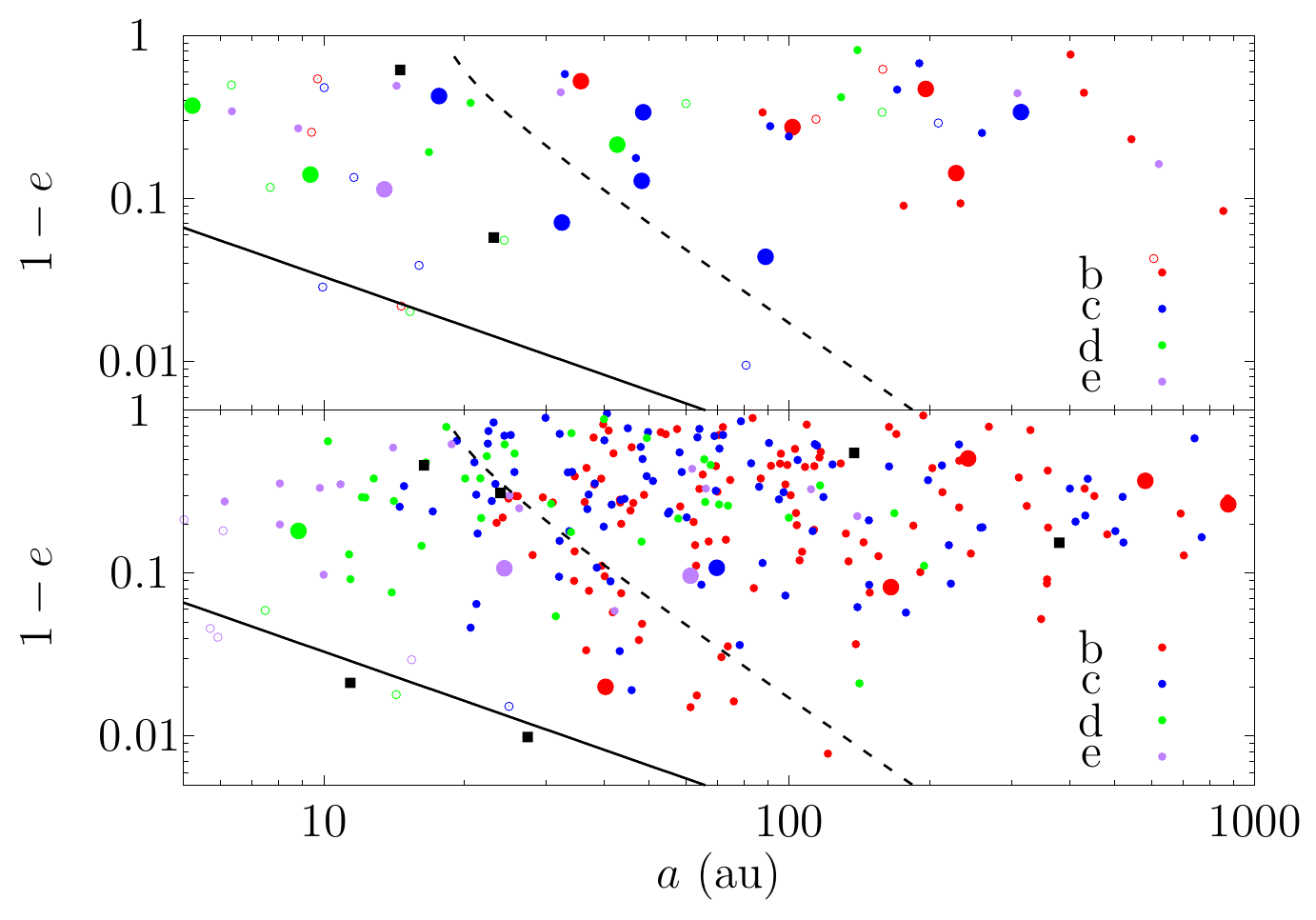}
\caption{Orbital distribution of captured planets from the HR 8799 system at the M-dwarf MD system. Bottom: lone captures where the M-dwarf captures only only planet; top: surviving captures where more than one is acquired during the encounter. Red, blue, green and purple represent HR 8799 b, c, d and e respectively. Filled-circles mean the M-dwarf system keeps the two super earths during both the encounter and the post-encounter evolution. Open-circles mean the system, though retaining the two super earths during the encounter, lose at least one later. Black squares represent the systems already losing at least one planet during the encounter. Big filled circles mean the two super earths in that system achieve a maximum relative inclination larger than 2 degrees during the post-encounter evolution. Captures below the solid line cross the orbit of the outer super earth at $0.33$ au and those below the dashed line may induce a precession rate in the outer planet' orbit faster than that by the coupling to the inner super earth.}
\label{fig-cap-orb-Mdwarf}
\end{figure}

We show in the bottom panel of Figure \ref{fig-cap-orb-Mdwarf} the orbits of the lone-captures. They mostly acquire very wide orbits from tens to hundreds of au and are often highly eccentric. A few of them may actually intersect with the orbits of the original close-in super earths. This is, however, very rare and extreme eccentricity is required. For example, a capture at 50 au needs to have $e>0.99$ to reach the inner planets. We use the solid line to show this limit, those below crossing the orbit of the outer original planet at 0.33 au; this happens for 5/288 of these lone-captures, thus with a chance of $\sim$1\%.

The remaining 99\% of the captured planets are most likely, interacting with the close-in super earths in a secular sense. The capture may force the inner planets to go through large-amplitude oscillations in eccentricity and inclination, called the Kozai--Lidov cycle \citep{Kozai1962,Lidov1962,Naoz2013}. If this cycle is activated, it may disrupt the super earth system (an example, though for the Kepler-48 system, is shown in Figure \ref{fig-ae-KP48-HR8799}). The timescale of this cycle \citep{Kiseleva1998}, in a normalised form, is
\begin{equation}
\label{eq-kl}
\begin{aligned}
T_\mathrm{KL}&\approx{2P^2_\mathrm{cap}\over 3\pi P}(1-e^2_\mathrm{cap})^{3/2} {m_\mathrm{cap}+m+m_\mathrm{*}\over m_\mathrm{cap}}\\
&\approx16\mathrm{Myr}\times{(a_\mathrm{cap}/100\mathrm{au})^3\over\left(a/0.33\mathrm{au}\right)^{3/2}}{\sqrt{m_\mathrm{*}/0.5m_\odot}\over m_\mathrm{cap}/5 m_\mathrm{J} } 
(1-e_\mathrm{cap}^2)^{3/2},
\end{aligned}
\end{equation}
where $P_\mathrm{cap}$ and $P$ are the orbital periods of the capture and of the original; $e_\mathrm{cap}$ is the orbital eccentricity of the capture; $m_\mathrm{cap}$, $m$ and $m_*$ are the masses of the capture, the original and the M-dwarf; $a_\mathrm{cap}$ and $a$ are the semimajor axes of the capture and the original; $m_\mathrm{\odot}$ and $m_\mathrm{J}$ are the masses of the Sun and Jupiter. So this timescale is typically $\sim 10$ Myr  in the case of a M-dwarf close-in planetary system and depends on $a_\mathrm{cap}$ to the third power.

The two original super earths themselves are also interacting with each other, forcing the orbits to precess and the eccentricity and inclination to oscillate with small amplitudes; this can be described by the Laplace--Lagrange secular theory (omitting the influence of the MMR). Following \citet{Murray1999}, the nodal precession timescale for the MD system is $T_\mathrm{LL}\sim 10^4$ yr. We note this is only a rough estimate because the proximity to/residence in MMR would have also contributed to the precession \citep[e.g.,][]{Callegari2004,Batygin2013a}.

The Kozai--Lidov cycle will be suppressed if its timescale is longer than that of the Laplace--Lagrange theory \citep{Innanen1997,Takeda2008,Lai2016,Mustill2017}. The dashed line in Figure \ref{fig-cap-orb-Mdwarf} denotes a line of constant $T_\mathrm{KL}$, equal to $T_\mathrm{LL}=10^4$ yr. Only captures under the line may induce large amplitude oscillations in $e$ and $i$ of the super earths through Kozai--Lidov cycles. For example, if a 5 Jupiter-mass planet (as assumed for the HR 8799 planets in this work) is captured at 50 au, its eccentricity has to be $\gtrsim0.93$. Compared to that of direct scattering, this requirement of $T_\mathrm{KL}<T_\mathrm{LL}$ is somehow less stringent and clearly more captures satisfy this criterion. But does the fulfilment of this criterion assure the system instability?

As shown in the bottom panel of Figure \ref{fig-cap-orb-Mdwarf}, very few captures, as marked as open circles, lead to loss of original planets. In all such cases, the captured planets are actually well below the dashed line, implying that $T_\mathrm{KL}<T_\mathrm{LL}$ criterion is too loose and should be treated with caution.

All the systems denoted by filled circles remain stable, not losing any of the two original planets and not only so, actually they mostly remain dynamically cold. We have recorded the relative inclination between the two super earths and use the large filled circles to represent systems where this angle ever becomes larger than $2^\circ$ in the post-encounter evolution -- this happens in only 9/288 systems. Further examination reveals that in 7 of them, the inclination is indeed caused by extremely close encounters $<3$ au (note the allowed range is 0-300 au in this set of simulations). Thus in only two instances, the captured planets manage to excite the mutual inclination between the two original planets; we confirm that the captures in both cases are below the dashed line.

When more than one planets from the HR 8799 system are captured by the M-dwarf, they may themselves scatter strongly. Here, among the 93 such captures, 67 survive after the post-encounter phase simulation. The top panel of Figure \ref{fig-cap-orb-Mdwarf} shows the orbits of the surviving captures and the symbol size/shape shows the properties of the original super earths of the M-dwarf system. Now significantly more M-dwarf systems are destabilised in the post-encounter evolution. And, the destabilisation does not depend on the orbits of the captures as stringently as that for lone captures.

These close-in M-dwarf systems seem fairly immune to captured planets from another planetary system. In the simulations, we use the HR 8799 system as the donor of the captured planets. For such wide donor planetary systems, capture is easier. But the captured planets often gain wide orbits, making it difficult to reach the original close-in super earths around the M-dwarf. Then a question arises: what if the M-dwarf close-in system encounters another system with intermediately-placed planets? This being the case, while the capture itself is harder, the captures, now more likely on smaller orbits, may be able to disturb the original planets more efficiently.

In order to assess this question while introducing as few changes as possible, we introduce a scaled-down version of the HR 8799 system. This is done by reducing the position and velocity vector of each planet with respect to the host star such that the outermost planet is moved to a stellar-centric distance of 12 au (this is where the innermost planet initially is) so the scale of the new system is $\sim$1/6 of the original. The masses of the objects are not touched and neither are the encounter parameters. Therefore, as viewed from the M-dwarf system, the only change is that the incoming planetary system is tighter. As such, the two sets of simulations can be directly compared. A total of 10000 runs for the encounter phase are carried out and the planet capturing systems are propagated through the post-encounter phase.
 
We observe that 60 planets from the scaled-down HR 8799 system are captured into 59 M-dwarf systems. As expected, the capture rate drop by 80\% compared those with the original HR 8799 system and the captured orbits are tighter here. During the post-encounter evolution, 12 of these M-dwarf systems lose at least one super earth and for 9 with both planets surviving, the relative inclination between the two becomes larger than $2^\circ$. The respective numbers from encounters with the original HR 8799 system are 26 and 20. Hence, planets on wider orbits seem to be more disruptive. A closer inspection reveals that it is mainly the multi-captures that make the difference: when more than one massive planet is captured, the mutual forcing between the giants makes them scatter, leading to the super earths' ejection/excitation. Because the original HR 8799 system is wider, 43 M-dwarf systems capture at least two planets whereas when encountering the tighter scaled-down HR 8799 system, only 1 M-dwarf system manages to do so.

\subsection{Kepler-48 system: flyby by a single Sun}

We have shown that the close-in planets are resistant to the encounter flybys. But what if the perturbed system has a planet at a somewhat more distant stellar-centric distance. In the Kepler-48 system for example, in addition to three inner planets inside 0.3 au there exists a gas giant at 2 au.

Before proceeding, we first note that the innermost planet Kepler-48 b, with an orbital period shorter than 5 days, may be subject to strong general relativity effect, causing fast orbital precession and potentially suppressing, e.g., Kozai--Lidov cycles \citep{Ford2000}. However, the short orbital period means that the other dynamical timescales could be short as well. Here we compare those of general relativity and the Laplace--Lagrange theory; both are competing with that of Kozai--Lidov cycles. It turns out that the innermost orbit is precessing on timescales of hundreds of years \citep[][]{Murray1999} and is much shorter than that of general relativity \citep{Naoz2013a}. Hence, the latter can be omitted and Newtonian dynamics suffice.

\begin{figure}
\includegraphics[width=\columnwidth]{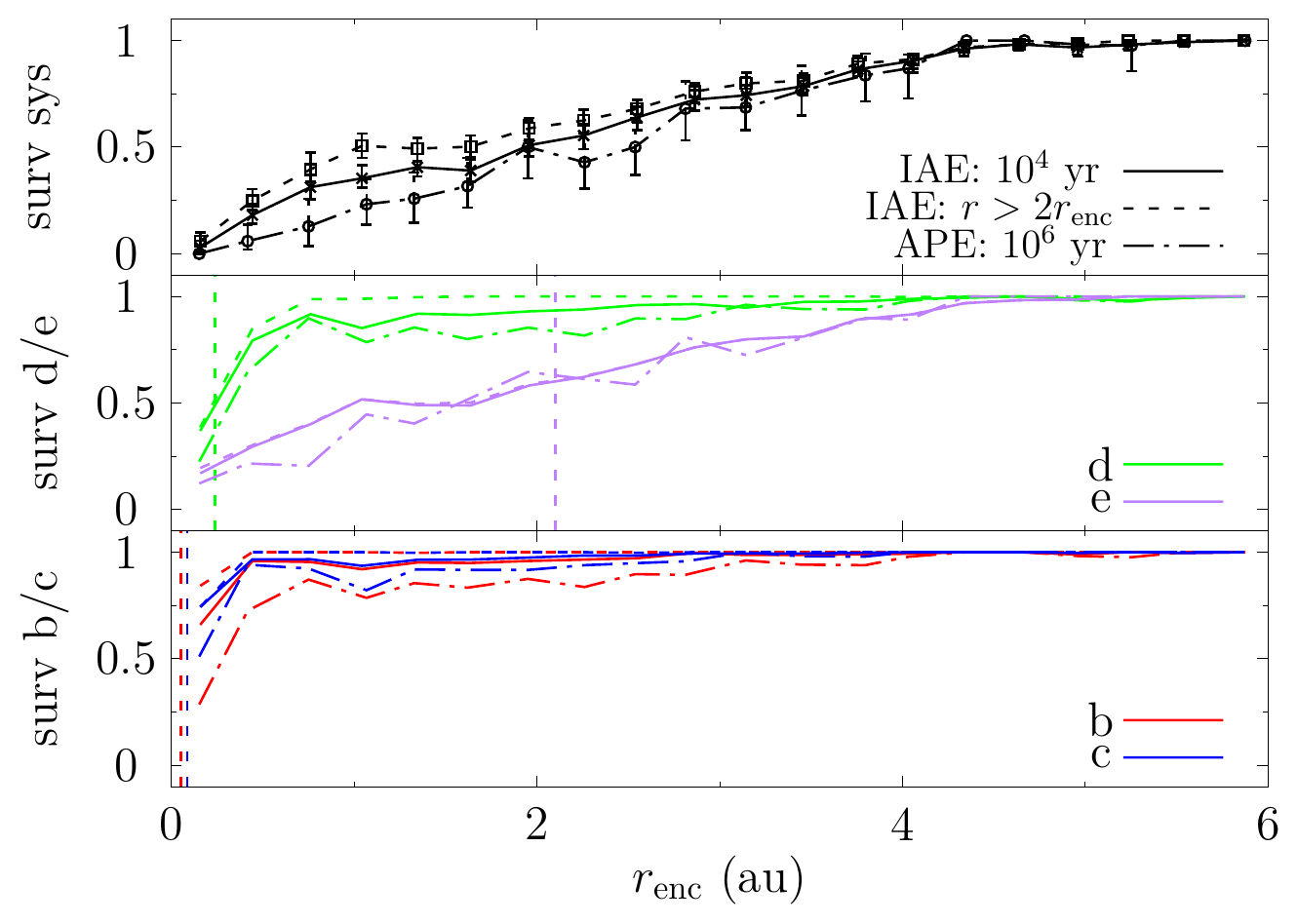}
\caption{Survival fraction of the Kepler-48 systems after encountering a single Sun at different times. The bottom two panels shows the planets in different colours; the vertical dashed lines mark the initial location of the planets. The top panel show the stability of the system as a whole. The dashed lines are for immediately after the encounter when the distance between the two stars are larger than twice the encounter distance (IAE: $r>2r_\mathrm{enc}$), the solid at $10^4$ yr (IAE: $10^4$) and dash-dotted for after the post-encounter evolution (APE: $10^6$).}
\label{fig-stab-kep48-sun}
\end{figure}

In bottom two panels of Figure \ref{fig-stab-kep48-sun}, we show the survivability of the planets at the two times in different colours. The solid lines are for the encounter phase (IAE: $10^4$ yr) and the dash-dotted lines for post-encounter phase (APE: $10^6$ yr). However, the short dynamical timescales suggest that our fixed integration time of $10^4$ yr for the encounter phase may be too long as significant ``post-encounter'' interplanetary interaction could have already taken place. Here we in addition record the system status measured at the point when the stellar mutual distance has passed the minimum and become larger than twice that value (IAE: $r>2r_\mathrm{enc}$), shown as dashed lines. At IAE: $r>2r_\mathrm{enc}$, as expected, the survivability of a planet only relies on its stellar-centric distance. Already at IAE: $10^4$ yr, the interplanetary forcing has led to the innermost also the least massive planet b to be damaged to a greater extent than its outer more massive neighbour c; and, d, the inner sibling of the outermost massive gas giant e experiences further loss at larger encounter distances. But overall, the three close-in super earths are resistant to the encounter, and only encounters inside 2 au can be detrimental to them. Different from those MD/MDR systems, here there is moderate post-encounter evolution in the sense that more planets are destabilised during that phase. This probably results from the combination of the effect of the more closely-packed and more massive inner planets themselves and the contribution from the outer gas giant Kepler-48 e. That planet itself is more susceptible due to its larger stellar-centric distance. Just like Jupiter in the solar system \citep{Hao2013,Li2019}, it is indestructible in the post-encounter phase owing to its dominance in the systems's mass budget. The top panel of Figure \ref{fig-stab-kep48-sun} shows the stability of the system. We observe that encounters out to 5 au can render the system unstable and overall, the post-encounter evolution is not significant.

\subsection{Kepler-48 system vs HR 8799 system: effect of captured planets}
We have seen in Section \ref{sec-mdwarf-hr8799} that for close-in M-dwarf planetary systems, a captured massive planet does not contribute to their system instability much, reason being that these tight super earths are well coupled and to break the coupling, the capture must acquire an extremely elongated orbit.

Compared to our M-dwarf systems, a key difference here is Kepler-48 e, the gas giant at a moderate stellar-centric distance of 2 au, relatively isolated from the inner super earth systems extending to 0.3 au. The Laplace--Lagrange theory predicts that the precession timescale of e due to the inner super earths is $\sim6\times10^4$ yr, again ignoring MMRs. Kepler-48 e is also disturbed by the captured planet. However, the adopted mass for Kepler-48 e is 2 Jupiter masses and is within a factor of a few compared to that of a capture (always 5 Jupiter masses in our simulations), so probably a general hierarchal three body problem model should be used \citep[e.g.,][]{Harrington1968}. Nonetheless, for simplicity and consistency, we still use Equation \eqref{eq-kl} to estimate the timescale of the precession in Kepler-48 e's orbit due to the perturbation of a capture.

\begin{figure}
\includegraphics[width=\columnwidth]{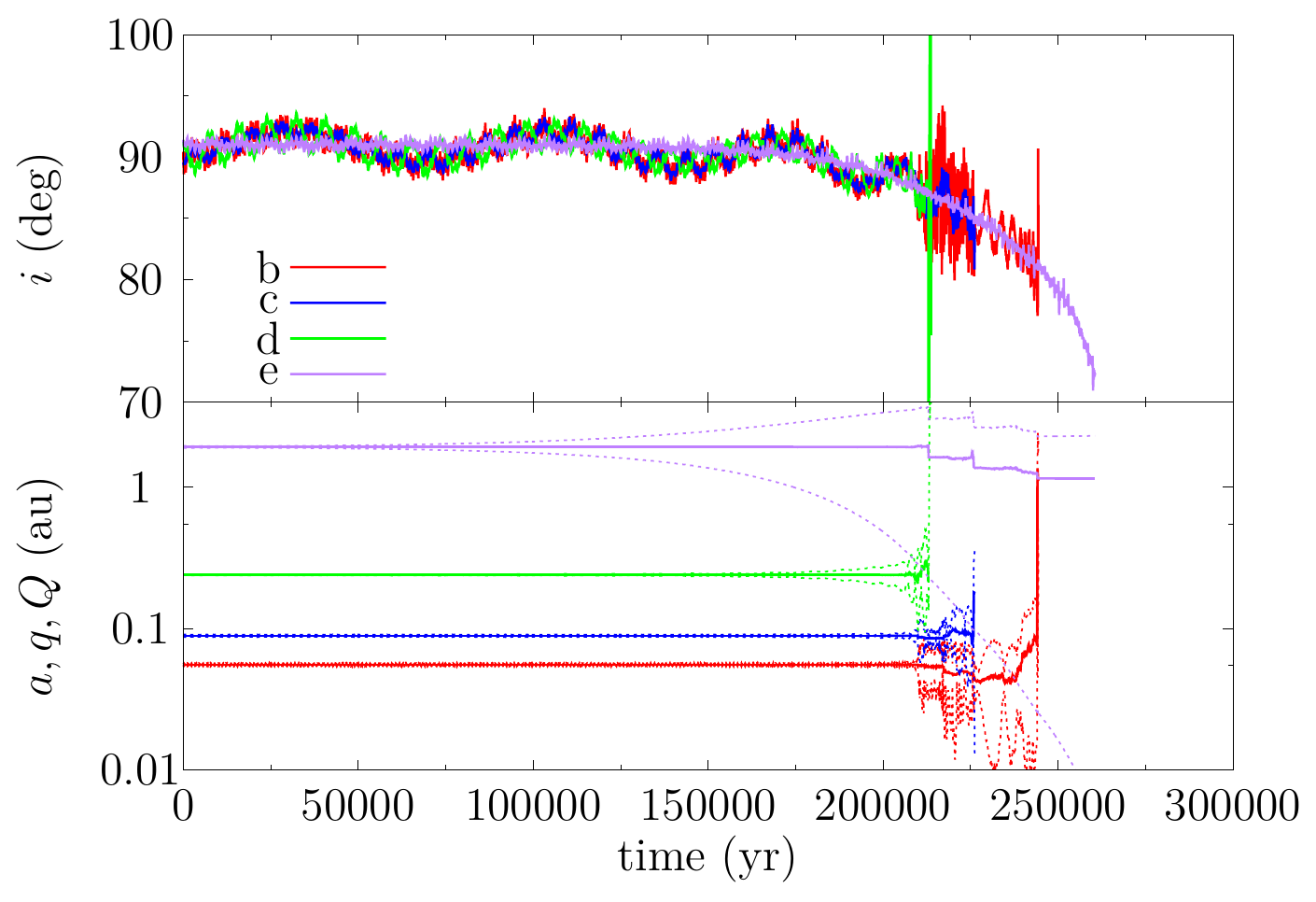}
\caption{Time evolution of orbital elements of a Kepler-48 system that captures a planet from the HR8799 system and remain intact during the encounter but later turns unstable in the post-encounter evolution. Bottom: that for $a$ (solid line), and pericentre $q$ and apocentre $Q$ distances (the two bracketing dotted lines) of the original four planets Kepler-48 b (red), c (blue), d (green) and e (purple). Top: that for $i$ of the four planets, all measured against the orbital plane of the capture. In this case, the capture has an orbit of $a=22$ au, $e=0.5$ and has been HR8799 c.}
\label{fig-ae-KP48-HR8799}
\end{figure}

Before the statistics, we first in Figure \ref{fig-ae-KP48-HR8799}, show an example system that is not disturbed much during the encounter but later becomes unstable during the post-encounter evolution, owing to a captured planet from the HR 8799 system. The bottom panel shows the temporal evolution of $a$ (solid), and pericentre $q$ and apocentre $Q$ (dotted) of the four original planets in different colours; the top panel depicts that of the orbital inclinations of the originals, all measured against the orbital plane of the capture, owing to its dominance in the total angular momentum of the entire system. In this instance, the captured planet has an orbit of $a=22$ au and $e=0.5$ is highly inclined, almost perpendicular to those of the originals (top panel of Figure \ref{fig-ae-KP48-HR8799}), a natural consequence of the isotropy of the captured orbits \citepalias{Li2019} that leads to a preference for high inclinations.

Kepler-48 e, the outer gas giant, is driven by the capture into large amplitude Kozai--Lidov cycles, as demonstrated by a steady increase of its $e$ accompanied by a decrease in $i$. In the meantime, the orbits of the inner three super earths experience no obvious variations and notably, their orbital planes tightly hold together. Once eccentric, e is able to force the eccentricity of its inner less massive neighbour d, probably through secular interactions, as $a$ is constant. When both orbits of Kepler-48 e and d become eccentric enough, they turn orbital crossing, as indicated by the fact that $Q$ of the inner intersects with the $q$ of the outer. From this moment onward, the direct scattering induces large variations in $a$, $e$ and $i$ of Kepler-48 d, making it encounter its inner siblings Kepler-48 c and then b; also the coplanarity is broken now. Soon, all three super earths are lost because of the scattering. Meanwhile, the eccentricity of e continues to increase to near unity \citep[e.g.,][]{Naoz2013} until it finally collides with the host star.

Then how often do the captures affect the stability of the Kepler-48 system? Out of the $10^4$ encounter phase simulations, 859 planets are captured from the HR 8799 system by Kepler-48. Among them, 537 are lone-captures, meaning that during the encounter, exactly one planet is captured by the Kepler-48 system; the remaining 322 are captured into 140 systems, each acquiring at least two.

\begin{figure}
\includegraphics[width=\columnwidth]{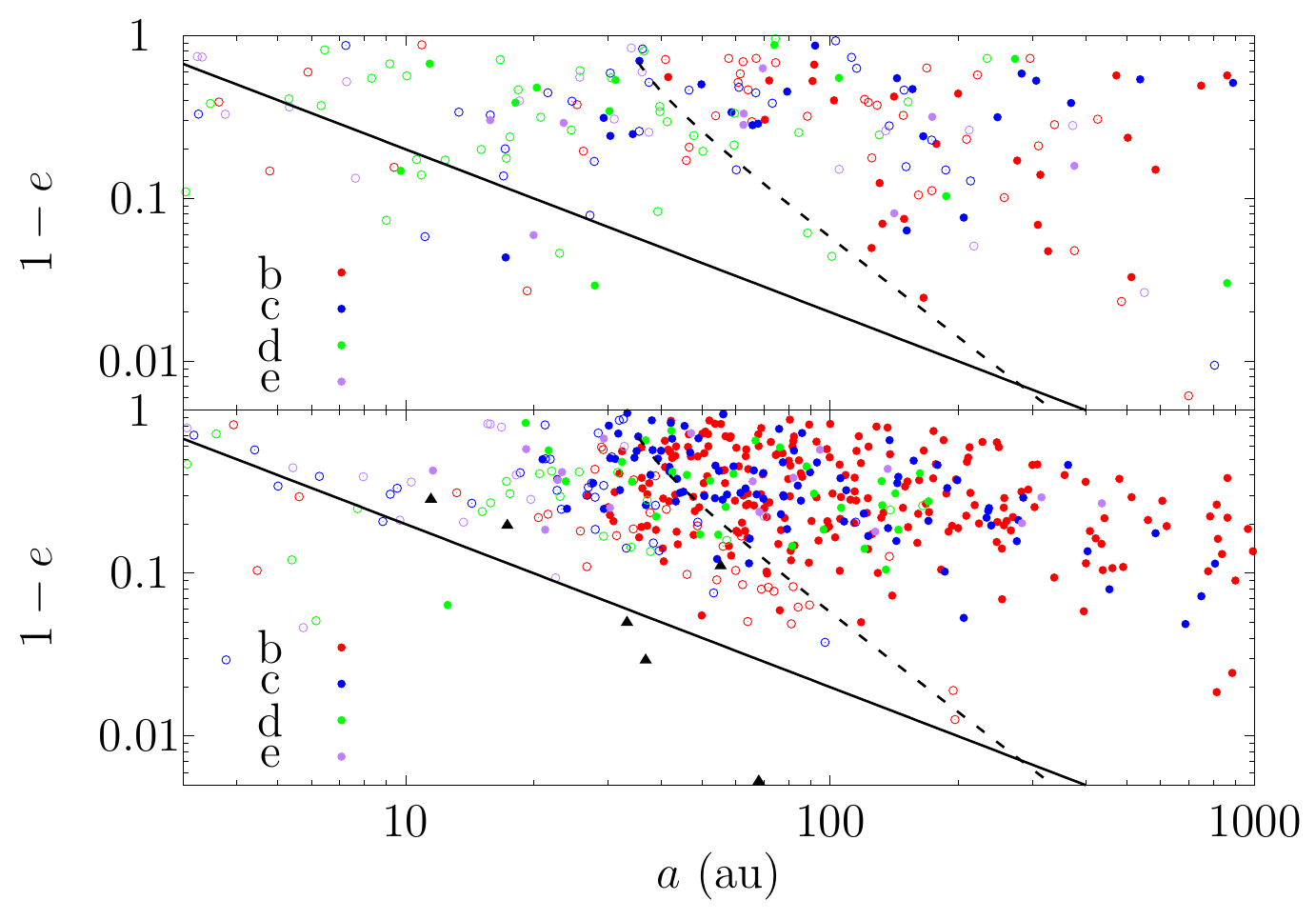}
\caption{Orbital distribution of captured planets from the HR 8799 system at the Kepler-48 system. Bottom: that for systems where only one planet is captured; top: that when more than one planets are captured. Red, blue, green and purple represent HR 8799 b, c, d and e respectively. Filled-circles mean the original planets in the Kepler-48 system remain stable during the post-encounter evolution irrespective whether or not in the earlier phase. Open-circles mean the system loses original planets in the post-encounter evolution. Captures below the solid line cross the orbit or the outermost original plant Kepler-48 e and those below the dashed line may induce a precession rate in the outer planet' orbit faster than that by the inner super earths. The black triangles represent captures that are soon ejected after being acquired before the end of the encounter phase simulation of $10^4$ yr.} 
\label{fig-cap-orb-KP48}
\end{figure}

In the bottom panel of Figure \ref{fig-cap-orb-KP48}, we show the orbits of all lone-captures. A complexity is that the capture, be it HR 8799 b, c, d or e, is always 5 Jupiter masses whereas the most massive original Kepler-48 e, is 2 Jupiter masses (cf., Table \ref{tab-sys}). Thus the capture itself may experience notable orbital evolution or even loss during the post-encounter phase, due to its interaction with the originals. Hence, plotted here are the orbital elements of the surviving captures after the post-encounter evolution.The solid line denotes the direct scattering limit with Kepler-48 e (so this is an equal-$q$ curve of $q=2$ au) and the dashed line marks the locations where $T_\mathrm{KL}=T_\mathrm{LL}$: captures below the solid line dip down to the regime of the original planets while those under the dashed line may force the orbit of Kepler-48 e to oscillate with large amplitudes. The unfilled circles mean that the system containing the capture lose {\it original} planets in the post-encounter evolution and filled points are used otherwise (the survivability of the system during the encounter is not considered here).

First we note that the orbits of the captures are similar to those in the M-dwarf systems (cf. Figure \ref{fig-cap-orb-Mdwarf}). Here, Kepler-48 e is only loosely linked to the inner super earths (as exemplified by the much longer $T_\mathrm{LL}$), and is thus more susceptible to the Kozai--Lidov cycles by the capture. And once the Kozai--Lidov cycles in e's orbit is activated, e can then disturb the inner super earth -- forming a chain-like phenomenon (see Figure \ref{fig-ae-KP48-HR8799} for an example). As a consequence, the Kepler-48 system is easier to disrupt. As indicated in the bottom panel of Figure \ref{fig-cap-orb-KP48}, here the $T_\mathrm{KL}=T_\mathrm{LL}$ level curve (dashed-line) is a fairly good stability indicator -- most of the systems below have lost at least one planet during the post-encounter evolution while most of those on the right do remain stable.

Notably, one of the lone captures lies below the solid line -- direct scattering with the originals is allowed -- yet they still remain stable in the post-encounter phase. The reason is that the capture itself is ejected. With a no significant mass hierarchy, the ejection of a capture is of no wonder and occurs in 36/537. The same could happen even before the end of the encounter phase simulation of $10^4$ yr. The black triangles show the orbits upon capture of the those planet that are are ejected already $10^4$ yr. Most of these lie around the solid line and are subject to strong scattering with the originals.

For systems capturing more than one planet, the final orbits for the surviving captures in the top panel of Figure \ref{fig-cap-orb-KP48} (215 planets in 127 systems). In most of the cases, the systems lose originals in the post-encounter phase. And the $T_\mathrm{KL}=T_\mathrm{LL}$ level curve is no good delimit between stable and unstable systems anymore.

\subsection{Implication}

We have shown that the close-in super earths around M-dwarf stars are rather resistant to flyby encounters in that such systems are hard to destabilise during both the encounter and the post-encounter phases. Only encounters closer than 1 au can possibly effect damage. And the mean motion resonance (MMR) between these does not play a role in the system's stability and the MMR itself is not much easier to break by the encounter.


The encounters between the M-dwarf systems and the HR 8799 planetary system enable the M-dwarf to capture planets from the latter at hundreds of au. In this way, the original super earths around the M-dwarf are not affected by the encounter much, allowing us to isolate the effect of the captures. However, the originals are so tightly coupled that very few captures with extremely small pericentre distance can decouple the originals, making them unstable or mutually-inclined. 

While the mutual inclination between the two original super earths remain, by and large, low (Figure \ref{fig-cap-orb-Mdwarf}), the two can be tilted together with respect to the initial reference plane (inducing large obliquity) which, though not rigorously defined, is probably a fair proxy of the equator of the central host. Here we only consider the intact systems where both originals survive the post-encounter simulation. At the end of the simulation, 22\% gain an obliquity larger than $20^\circ$ and 6\% become even retrograde. These are probably lower limits because our post-encounter integration of 1 Myr could be only a fraction of the Kozai--Lidov timescale \eqref{eq-kl} and thus when the simulation is stoped, the obliquity may be still increasing.

Hence, with the perturbation of an outer massive planets, we obtain ``cold'' multi-planet systems that are, while coplanar with respect to each other, inclined against the spin axis of the host star. This reminds us of the Kepler-56 system. There, a Neptune-massed and a Saturn-massed planet are orbiting a red giant in a mutual-coplanar configuration, both with orbital periods of a few tens of days. The intriguing feature is that the planetary orbits are misaligned with the spin of the central host, by tens of degrees \citep{Huber2013,Li2014b}. In that system, there also exists a third, 5-Jupiter mass planet on 1000-day orbit \citep{Otor2016}. Our result potentially raises the possibility that such systems have an external cause \citep[see also][]{Li2014b,Gratia2017}.

Compared to the M-dwarf system described above, the Kepler-48 system, due to the existence of the gas giant at 2 au, are easier to destabilise. And if the host star captures additional planets on wide orbits, because this giant planet is not strongly coupled to the inner system, it can be driven into large amplitude Kozai--Lidov cycles, leading to intense scattering among the inner less massive siblings.


\section{Planets on intermediately placed orbits} \label{sec-solar}
Our intermediately place planets are less close-in with outer planets reaching a few tens of au. Thus, they should be more vulnerable to encounter flybys.

\subsection{Three-Jupiter systems: flyby by a single Sun}
We begin with our 3J10 and 3J15 simulations. In Table \ref{tab-3J} we show the fraction of systems that remain stable at the two phases, categorised by the encounter distance.

During encounters inside 33 au, or equivalently, closer than the stellar-centric distance of the outermost planet, 30\% of the 3J10 systems are able to keep all planets. The survival fraction for 3J15 during the encounter is similar, a result of the same stellar-centric distances of outermost planets in the two systems. During the post-encounter phase, the evolution of the two systems diverge as the planets' mutual separations affect the timing of the instability. As such, at 100 Myr, only 7\% of the 3J10 system remain stable after an encounter inside 33 au while this is 20\% for the 3J15 simulations. This feature -- agreement during the encounter and disagreement during the post-encounter evolution -- is also the case for other encounter distance ranges.

\begin{table}
\centering
\caption{Fraction of systems remaining stable at the encounter and the post-encounter phases for the 3J10 and 3J15 systems after encountering a single Sun.}
\label{tab-3J}
\begin{tabular}{c c c c c}
\hline
\multirow{2}{*}{$r_\mathrm{enc}$ (au) } &  \multicolumn{2}{c}{3J10}& \multicolumn{2}{c}{3J15} \\
  & encounter &post-encounter&encounter &post-encounter\\
\hline
0-33 & $0.30_{-0.04}^{+0.05}$ & $0.07_{-0.02}^{+0.03}$ & $0.31_{-0.05}^{+0.06}$&$0.19_{-0.03}^{+0.05}$\\
33-67 & $0.84_{-0.04}^{+0.04}$ & $0.51_{-0.06}^{+0.05}$ &$0.80_{-0.05}^{+0.04}$ &$0.65_{-0.06}^{+0.05}$\\
67-100 & $0.97_{-0.02}^{+0.01}$ & $0.81_{-0.04}^{+0.03}$ &$0.97_{-0.02}^{+0.01}$&$0.93_{-0.02}^{+0.03}$ \\
\hline
\end{tabular}
\end{table}

\begin{figure}
\includegraphics[width=\columnwidth]{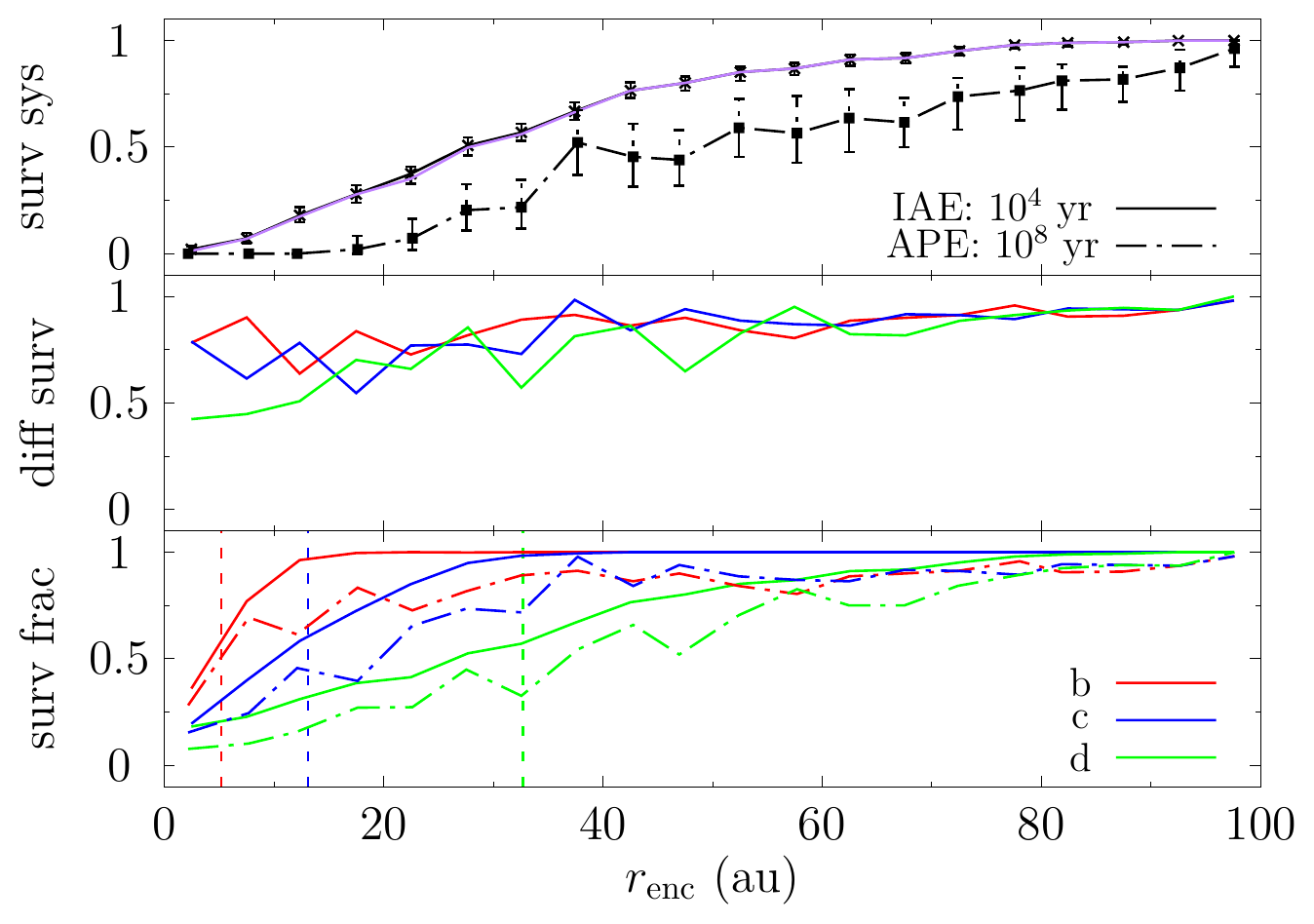}
\caption{Stability of the 3J10 systems during the encounter and post-encounter phases after a single Sun. Bottom: stability of the three planets; red, blue and green for the inner, middle and outer planets, respectively; solid lines are for the encounter phase (IAE) and dash-dotted for post-encounter (APE). Middle: fraction of planets surviving the encounter that also remain stable during the post-encounter phase. Top: fraction of systems that keep all three planets during the two phases (black) and the product of the individual probability that each of the three remain stable IAE (purple).}
\label{fig-stab-3j10}
\end{figure}

Figure \ref{fig-stab-3j10} shows the stability of the planets and the system as a function of the encounter distance at the two phases for the 3J10 systems. Presented in the bottom panel are those for the three planets in different colours, solid for the encounter and dash-dotted for the post-encounter phase. In both phases, the rates clearly depend on the position of the planets: outer planets are easier to destabilise. However, it seems that the loss of planet during the post-encounter evolution, i.e., the difference between the solid and dash-dotted lines, is similar for the three planets. The middle panel, showing the fractional loss during this later phase, clearly supports this point. Thus, the initial location of a planet plays a minor role in the post-encounter phase and any planet, irrespective of its stellar-centric distance, is disrupted to a similar degree \citep[reminiscent of the classical planet scattering without stellar encounters][]{Marzari2002}. The top panel presents the survivability of the system as a whole. Here the black solid line denotes the fraction of systems that retain all three planets during the encounter and the solid purple line is simply the product of the rates for each of the three planets to remain bound. The agreement between the two indicates that the stability of each planet is unrelated during the encounter \citepalias{Li2019}. Later in the post-encounter evolution, more systems (the difference between the solid and the dash-dotted lines) are destabilised and this does not depend on the encounter distance distance much: e.g., an encounter at 30 au is no more disruptive than one at 60 au. At $r_\mathrm{enc}\approx95$ au, an encounter cannot induce system destruction anymore.

The case for 3J15 systems is similar but with higher fraction of survival and we omit the detailed discussions.

\begin{figure}
\includegraphics[width=\columnwidth]{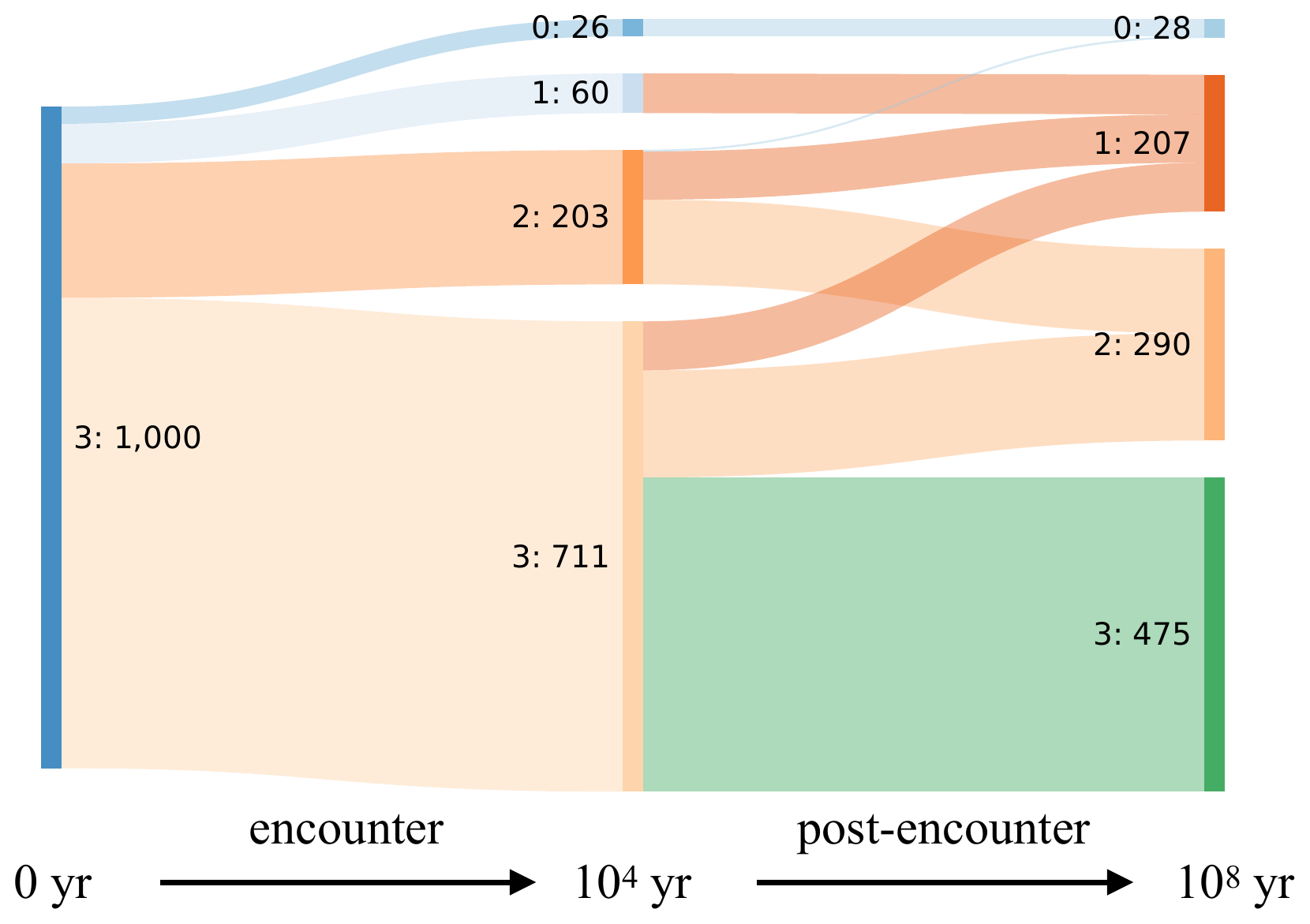}
\caption{Evolution of the number of planets in the 3J10 systems at different times after encountering a single Sun. On the left is the initial state before the encounter, in the middle that immediately after the encounter and on the right that after the post-encounter simulation. Each label has two number: the first is the number of planets in a system and the second the number of cases observed.}
\label{fig-system_state}
\end{figure}

Then in Figure \ref{fig-system_state} we report the evolution of the number of planets in a system for the 3J10 simulations. At the beginning and thus before the encounter, all have 3 planets. During the encounter, the majority, $\sim$700/1000, are able to retain all planets while the other 300 lose at least one planet; among these, 200 lose only one and 20 are deprived of all the three. During the post-encounter evolution, as indicated in Table \ref{tab-3J} and Figure \ref{fig-stab-3j10}, these systems undergo strong scattering and suffer from (further) loss of planets -- more than half of the systems are damaged. As a result of the two phases of evolution, two systems, can evolve differently but reach the same final state. For instance, a system can keep all planets during the encounter and lose one later (3->3->2) or it may lose one early during the encounter but preserve the remaining two during the post-encounter evolution (3->2->2). Can the two systems be somehow distinguished?

\begin{figure}
\includegraphics[width=\columnwidth]{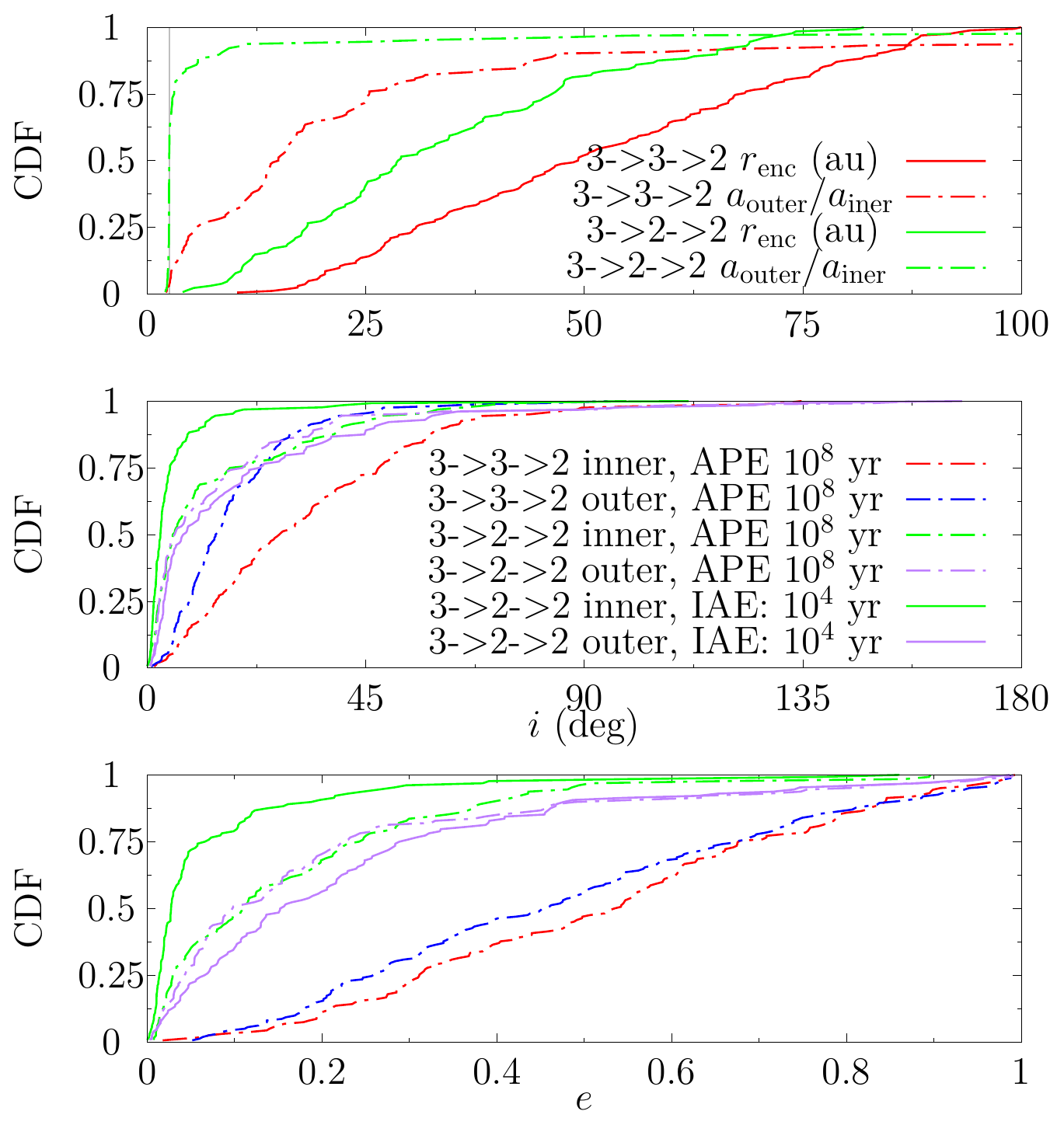}
\caption{Cumulative distribution function (CDF) of orbital elements of the 3J10 systems that are left two planets after the post-encounter simulation after encountering a single Sun. Bottom: CDFs for final $e$ of the inner (red) and outer (blue) planets in systems that keep all three during the encounter and lose one in the post-encounter evolution (3->3->2); CDFs for the inner (green) and outer (purple) planets in systems that lose one during the encounter and the remaining survive after the post-encounter evolution (3->2->2): solid and dash-dotted lines are for immediately after the encounter (IAE) and for after the post-encounter evolution (APE) respectively. Middle: the same as the bottom but for $i$. Top: CDF of the ratio of the semimajor axes of the outer to the inner survivor (dash-dotted) and  of the encounter distance (solid) for the 3->3->2 (red) and 3->2->2 (green) systems.}
\label{fig-aeicdf}
\end{figure}

In the bottom panel of Figure \ref{fig-aeicdf}, we show the cumulative distribution function (CDF) of the eccentricity of the two surviving planets (inner and outer) for the 3->3->2 and 3->2->2 systems. That for 3->3->2 is in red and blue for the inner and outer surviving planets, both in dash-dotted lines and for the final distribution after the post-encounter evolution. That for 3->2->2 is in green and purple for the inner and outer final survivors, dash-dotted for post-encounter evolution and, additionally, solid for immediately after the encounter. The two planets in the 3->3->2 systems are characterised by similar CDFs, the inner one slightly hotter, like that in pure planet scattering simulations \citep{Marzari2002,Chatterjee2008}, implying that this may be the main driver for eccentricity excitation. As for the 3->2->2, the outer planets are clearly perturbed more significantly during the encounter with a much hotter CDF. During the post-encounter phase, nonetheless, the difference is removed by the interplanetary forcing and both planets acquire the same CDF. Notably, systems via 3->3->2 is significantly hotter than that for 3->2->2. For example, in the former, about half of the planets have $e>0.5$ and for the latter, only 10\% of the orbits become this eccentric. This is probably because the encounter can liberate the outermost planets without perturbing the inner two planets too much and such systems more or less evolve secularly.

The middle panel of Figure \ref{fig-aeicdf} shows the CDF for inclination and the result is similar to that of $e$. For instance, the 3->2->2 systems have hotter outer planets and colder inner ones immediately after encounter but this characteristics is wiped out during the post-encounter phase, both planets reaching a similar CDF. Also, just like the classic planet scattering simulations \citep{Marzari2002}, for the 3->3->2 systems, the outer planet have a smaller inclination because of the large semimajor axis and consequently, a larger amount of ``inertia'' to overcome when tilting its orbits.

In the top panel, we show the CDF for the final ratio of semimajor axes of the two surviving planets $a_\mathrm{outer}/a_\mathrm{inner}$ and the $r_\mathrm{enc}$ of the encounter leading to such systems. The red/green dash-dotted lines show $a_\mathrm{outer}/a_\mathrm{inner}$ for the 3->3->2 and 3->2->2 systems after the post-encounter evolution. While most of the 3->2->2 systems have $a_\mathrm{outer}/a_\mathrm{inner}$ around the initial value (grey vertical line), the 3->3->2 systems is characterised by much larger values and indeed 2/3 of them have $a_\mathrm{outer}/a_\mathrm{inner}>10$, forming highly hierarchical configurations \citep[cf., e.g.,][]{Marzari2002}. Finally, we read from the solid lines that encounters responsible for the 3->3->2 systems are in general much wider than those for the 3->2->2 ones, as all three planets are kept during the encounter.

One may wonder that whether or not the 3->3->2 systems are genuine in that they may become already orbital crossing during the encounter phase. Actually about half of these systems present this behaviour. But their final distribution of orbits is indistinguishable from those not orbital-crossing. So both are presented together in Figure \ref{fig-aeicdf}.

\subsection{NUSJ: flyby by a single Sun}
\begin{figure}
\includegraphics[width=\columnwidth]{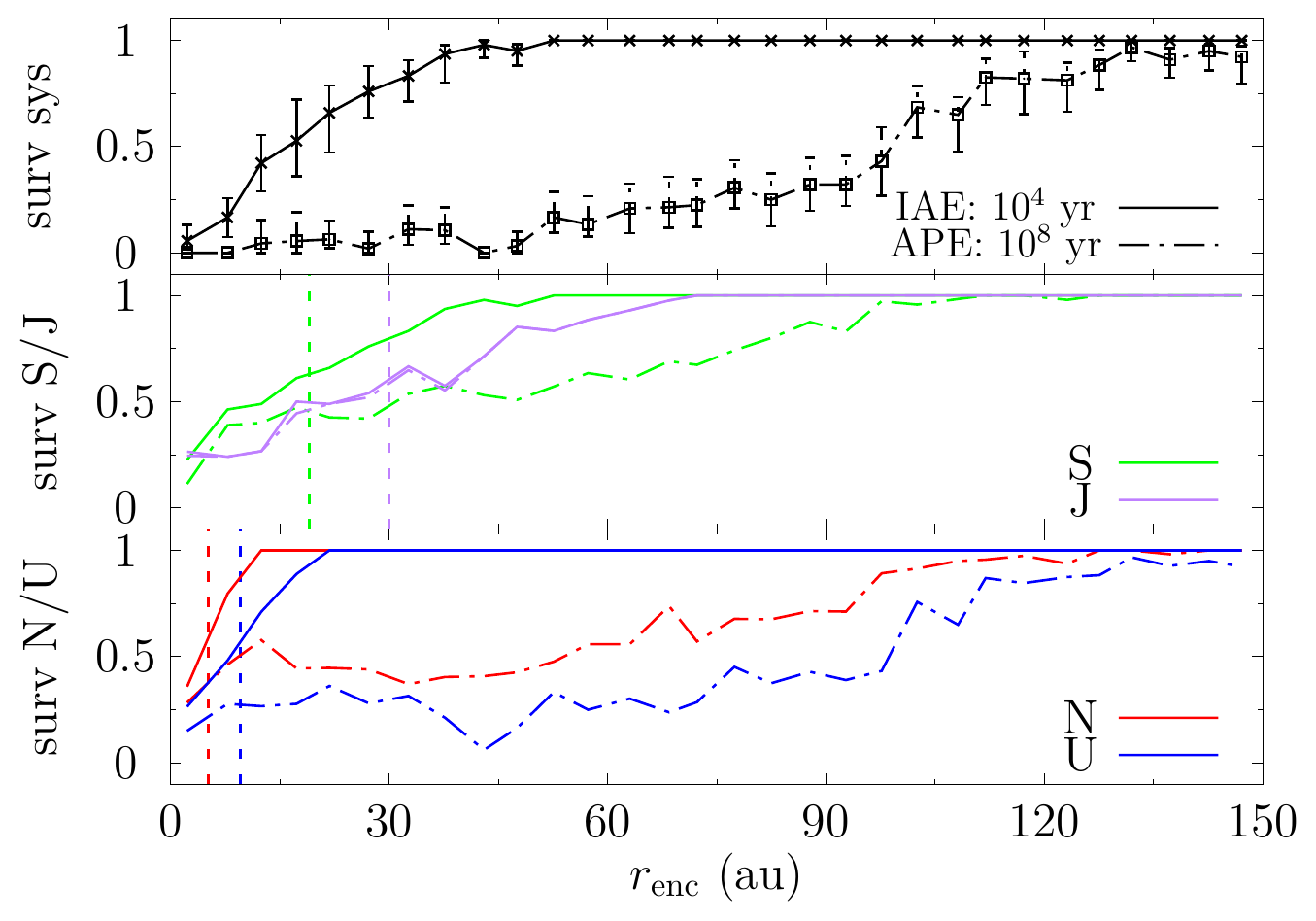}
\caption{Survival fraction of the planets in the NUSJ systems as a function of the encounter distance after encountering a single Sun. Those for the ice giants are in the bottom panel, red for Neptune and blue for Uranus; In the top panel: green is for Saturn and purple is for Jupiter. Solids line are for immediately after encounter and dash-dotted for after the post-encounter simulation.}
\label{fig-stab-NUSJ}
\end{figure}

The stability of the NUSJ simulations are presented in Figure \ref{fig-stab-NUSJ} as a function of the encounter distance. There, the solid and dash-dotted lines show the survival fraction of the four planets in different colours in the bottom two panels immediately after encounter (IAE) and after the post-encounter evolution (APE).

Jupiter, the outermost planet, is the most vulnerable during the encounter as expected. But owing to its large mass, it is almost immune to the post-encounter evolution. Saturn, while less ejected during the encounter, experiences much more loss later. The two ice planets, also the most resistant to the direct effect of the encounter, suffer from the most disruption during the post-encounter evolution because of their small masses.

We compare the results to \citetalias{Li2019} where the system was JSUN (solar system giant planets in the true order). As the encounter phase is prescribed mainly by the heliocentric distances of the planets, the results are similar in the two configurations. But clearly the mass budget plays a dominant role in the post-encounter evolution -- the most massive planets, Jupiters, are always the most stable while Uranus, the least massive, is lost the most frequently. So here the stellar-centric distances are essentially irrelevant \citep[but there is extreme hierarchy in the sizes of the orbit, inner smaller planet can eject outer larger ones; cf.][]{Mustill2015}.

Nonetheless, a major distinction from \citetalias{Li2019} appears that, while for a JSUN system, encounters at 100 au can hardly do any damage, a NUSJ system is susceptible to encounters even at 140 au. What makes the difference?

It turns out that the angular momentum deficit of a system \citep[AMD,][]{Laskar1997,Chambers2001}, defined as the amount of angular momentum needed to make all the planets' orbits circular and coplanar while keeping semimajor axes unchanged, appears be a key parameter. Quantitatively,
\begin{equation}
\label{eq-amd}
\text{ AMD}\approx\sum_{j} m_{j}\sqrt{a_{j}}\left(1-\sqrt{1-e^2_{j}}\cos{i_{j}}\right),
\end{equation}
where $m_j$ is the mass of the $j$th planet and $a_j$, $e_j$ and $i_j$ its orbital semimajor axis, eccentricity and inclination; the summation runs over all planets.  The AMD of a system is conserved in the secular evolution \citep{Laskar2017}. And, if the AMD is allowed to flow freely within a system and sufficiently large, orbital crossing will occur and such a system is called AMD-unstable; we call this critical value AMD$_\mathrm{cross}$ and it is obtained following \citet{Laskar2017}. We note that the AMD-unstable systems are not necessarily unstable \citep[e.g., the terrestrial planets in our own solar system is AMD-unstable but probably stable over Gyr timescales; see][]{Laskar1997}.

\begin{figure}
\includegraphics[width=\columnwidth]{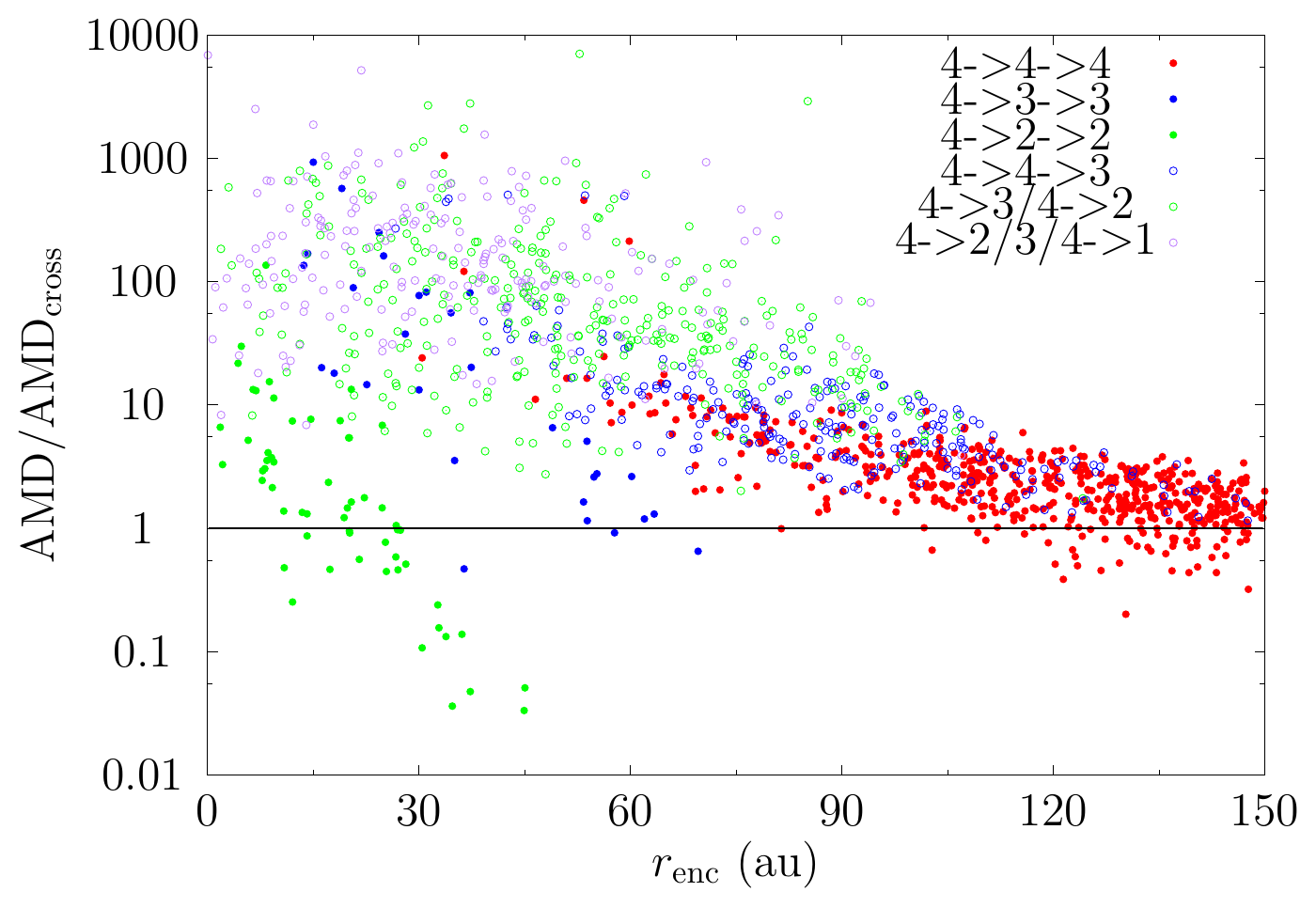}
\caption{Angular momentum deficit (AMD) of the NUSJ systems immediately after the encountering a single Sun normalised by AMD$_\mathrm{cross}$, the minimum needed for the systems to have orbital crossing. Different point types/colours represent different evolutionary tracks: in the figure legend, the first number is the original number of planets (all being 4), second that immediately after the encounter and the third after the post-encounter evolution. Filled points denote systems that remain stable in the post-encounter evolution and unfilled symbols otherwise. Red, blue, green and purple points are used to represent systems that, after the post-encounter evolution, have 4, 3, 2 and 1 planet. Systems retaining only one planet immediately after the encounter are omitted because their AMD$_\mathrm{cross}$ is not well defined.}
\label{fig-amd-stab-NUSJ}
\end{figure}

In Figure \ref{fig-amd-stab-NUSJ}, we show the AMD of the NUSJ systems with at least two planets {\it immediately after the encounter} as a function of the encounter distance. The AMD has been normalised with respect to AMD$_\mathrm{cross}$ for the respective system; systems with AMD over unity are AMD-unstable and -stable otherwise. To distinguish between AMD-stability and the stability of a system in our simulations, we call the latter ``actual-stability'' within this part of the paper. Filled circles represent actually-stable systems -- those remain intact during the post-encounter evolution (but perhaps lose planets already during the encounter). Red, blue and green points mean that the systems end with 4, 3 and 2 planets after the post-encounter evolution (and this is also the number of planets immediately after the encounter since the systems are actually stable). Unfilled circles denote those actual-unstable in the post-encounter evolution and are, at the end of this phase, left with 3 (blue), 2 (green) and 1 (purple) planet.

Two obvious observations are: 1) the AMD excitation, normalised with respect to the AMD$_\mathrm{cross}$ of each individual system, is clearly a function of the encounter distance and 2) there is a correlation between AMD- and actual-stability.

In a closer examination, we find that even during the most distant encounters, i.e., $r_\mathrm{enc}\gtrsim120$ au, most of these NUSJ systems keep four planets with AMDs a few times AMD$_\mathrm{cross}$; but the majority of them are actually stable. A small fraction, all AMD-unstable, lose a planet, mostly likely an ice giant (cf. Figure \ref{fig-stab-NUSJ}). Closer-in, around 90 au, the encounters are able to raise the system's AMD above AMD$_\mathrm{cross}$ by an order of magnitude; accordingly, half of the systems lose at least one planet during the post-encounter evolution. Inside of 60 au, the AMD immediately after the encounter is greater (some reaching AMD/AMD$_\mathrm{cross}\sim100$) and do turn actually unstable during the post-encounter evolution; but a few, often with small AMD/AMD$_\mathrm{cross}$ are capable of remaining actually stable during the post-encounter evolution. For the closest encounters inside 30 au, the scattering is large with AMD/AMD$_\mathrm{cross}$ reaching 10000. Nonetheless, still, a few three- or two-planet systems can be actually stable and their AMD/AMD$_\mathrm{cross}$ ranges between 0.1 and 1000.

Noteworthily, a few systems, though endowed with an AMD $\sim100$ times AMD$_\mathrm{cross}$ during the encounter, remain actually stable with all the four planets over the entire post-encounter phase of 100 Myr. This is because, the outermost planet Jupiter is emplaced by the encountering star onto a wide and eccentric orbit well decoupled from the inner system yet it carries an AMD orders of magnitude larger; on the other hand, the inner planets are not touched much by the encounter. Because of the hierarchy, this system is actually stable \citep[for instance,][]{Innanen1997,Takeda2008}. We confirm that this is also the case for the few systems left with three planets and huge AMD but actually stable.

\begin{figure}
\includegraphics[width=\columnwidth]{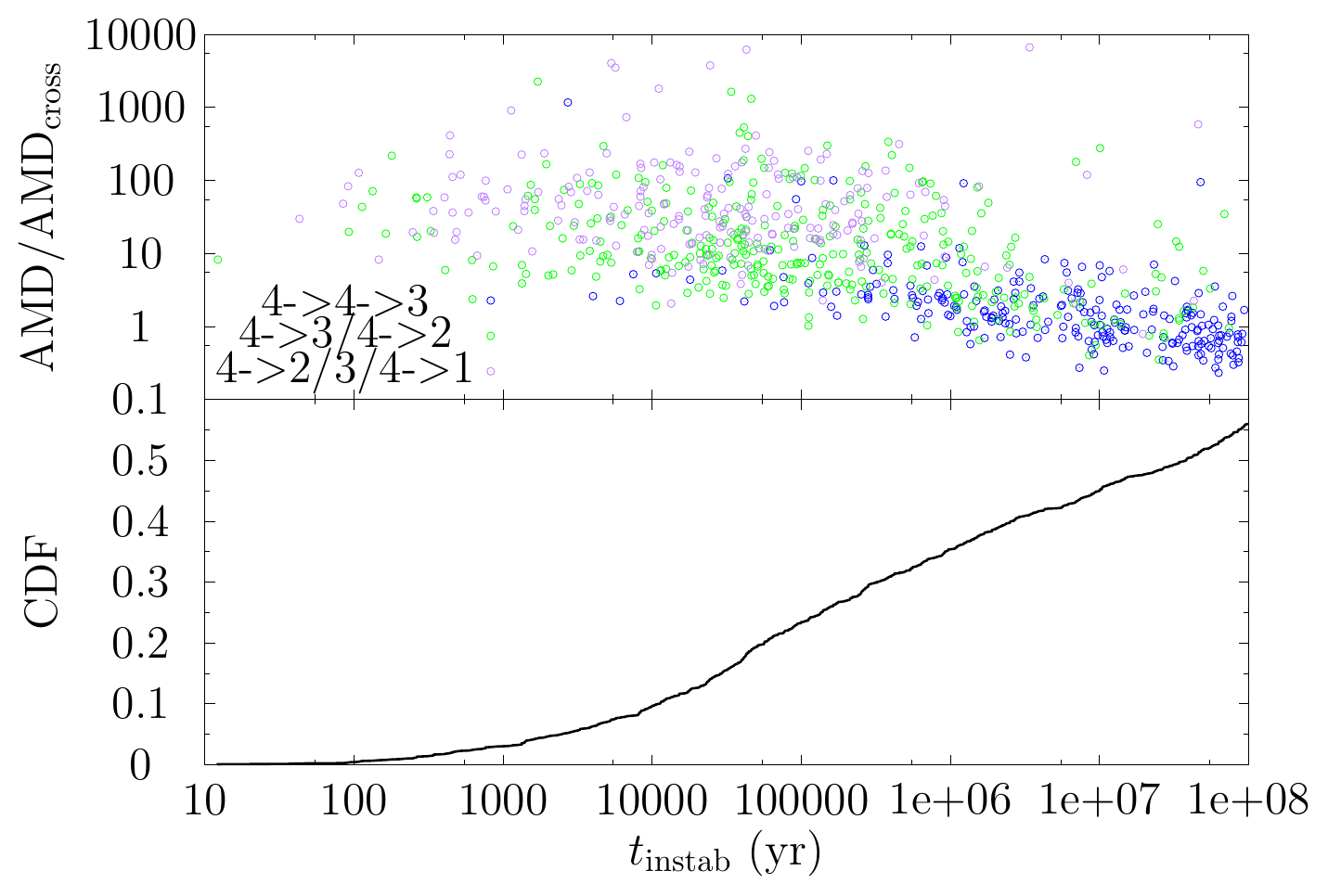}
\caption{The time of the first planetary encounter $t_\mathrm{instab}$ in the actually-unstable NUSJ systems after encountering a single Sun. Bottom: the cumulative distribution function of $t_\mathrm{instab}$. Top panel: AMD/AMD$_\mathrm{cross}$ of these systems as a function of $t_\mathrm{instab}$ immediately afte the encounter; the colours correspond to different numbers of surviving planets after the post-encounter evolution: blue for 3, green for 2 and purple for 1.}
\label{fig-instab-time-amd}
\end{figure}

Then, the fact that most of these actually-stable systems are AMD-unstable seem to suggest that they may become unstable, beyond our integration time of $10^8$ yr. An additional complexity is that in the original solar system, the outer two, Uranus and Neptune are close to 2:1 MMR \citep{Callegari2004}. When turned into NUSJ, the much larger masses of the two gas giants make the MMR stronger and may facilitate the transfer or even the creation of AMD \citep{Wu2011}. In the bottom panel of Figure \ref{fig-instab-time-amd}, we show the cumulative distribution function (CDF) of the times when the first planetary encounter (which we loosely define as the instant when any two planets' mutual distance becomes smaller than 0.5 au) occurs $t_\mathrm{instab}$ in an actually-unstable system. The CDF follows a straight line beyond $10^4$ yr into the post-encounter phase simulation. Hence, we probably miss out some later instability. In the top panel, the AMD (scaled with AMD$_\mathrm{cross}$) of these actually-unstable systems are presented as a function of the instability time $t_\mathrm{instab}$. Though with large scattering, there seems to be a linear correlation between $\log$~(AMD/AMD$_\mathrm{cross}$) and $\log t_\mathrm{instab}$. This implies that the role $\log$~(AMD/AMD$_\mathrm{cross}$) is playing is similar to that of the planets' separation $K$ \citep[as measured in the mutual Hill radius, see Equation \ref{eq-k} and cf. ][]{Chambers1996,Zhou2007}. Hence AMD/AMD$_\mathrm{cross}$ immediately after the encounter may be used to predict the longevity of the system in the post-encounter phase.

Finally, we come back to the question why NUSJ systems can be more easily disrupted by a more distant encounter than the JSUN counterparts. Given that the planetary interaction can be largely ignored during the encounter, two planets in the two types of systems, whatever their masses, if positioned the same, receive similar amounts of ``specific perturbation''. Hence, in the NUSJ simulations, it is the most massive planet Jupiter to which the largest specific nudge is imparted. This specific disturbance should be the same as that Neptune acquires in the JSUN system given a similar encounter. But because of the much larger mass of Jupiter than that of Neptune, the amount of AMD increase in the two systems are different and the NSUJ systems gather much more AMD than the JSUN equivalents. And probably, the difference in AMD explains the different dependence on the encounter distance.

To confirm this, we carry out a simple experiment by switching the planetary masses in the NUSJ systems to make them effectively JSUN (i.e., by assigning Jupiter's mass to Neptune, etc.). For these ``JSUN'' systems, we calculate their AMD and confirm that they do collect less AMD (scaled with AMD$_\mathrm{cross}$) during the encounter. We have also revisited the real JSUN simulations in \citetalias{Li2019} and agreement is found.

\subsection{Implication}
Here we have examined the influence of encounters on systems with three intermediately-placed Jupiters, either with planets' separations of 10 (3J10) or 15 (3J15) times their mutual Hill radii. While both systems suffer from loss of planets, the more compact ones (3J10) are clearly more vulnerable and provide better statistics on which thus we concentrate. Our result shows that during the encounter, a planet's stability crucially relies on its stellar-centric distance. Such a dependence is wiped out during the post-encounter evolution and all planets are equally susceptible.

As a result of the two phases of evolution, a system can, for instance, lose a planet during the encounter phase and remains stable later on, or it can keep all three in the first phase and lose one in the post-encounter phase. We have compared the the distribution of the surviving two planets' orbital elements in the two cases. Our result shows that those losing a planet in the post-encounter evolution have distributions resembling the classical planet scattering (without stellar encounter) and significantly hotter than those of the other evolutionary path.

We have also explored the evolution of the reversed solar system ``NUSJ'', to study the effect of the mass gradient. The stellar-centric distances determine the stability during the encounter but it is the planetary masses that matter the most for the later post-encounter evolution. Jupiter is almost immune to the post-encounter phase evolution while Uranus is lost to the largest degree.

We have calculated the angular momentum deficit (AMD) of these NUSJ systems immediately after the encounter. It appears that AMD directly relates to the encounter distance and to the stability in the later post-encounter phase. The closer the encounter distance, the larger the AMD and the system is more likely to be unstable (Figure \ref{fig-amd-stab-NUSJ}). Additionally, for these unstable systems, their AMD is correlated to the instability time -- the larger the AMD the earlier the instability (top panel of Figure \ref{fig-instab-time-amd}).

If inspecting Figures \ref{fig-amd-stab-NUSJ} and  \ref{fig-instab-time-amd}, we find that different symbols/colours (thus systems with different evolutionary tracks) seem, to some extent, separated from each other. A prominent example is a strip of filled green circles in Figure \ref{fig-amd-stab-NUSJ} (those left with two planets during the encounter and are stable in the post-encounter evolution). We have revisited the JSUN simulations in \citetalias{Li2019} and the 3J simulations in this work, all showing such segregations.

We comment that AMD is a single parameter encompassing the various information (Equation \eqref{eq-amd}), e.g., the planets' masses, stellar-centric distances and their eccentricities. The above observations show a system's AMD immediately after the encounter closely relates to the system later post-encounter evolution, in determining both the stability and the the instability times. This implies maybe AMD can be used to predict the fate of a planetary system owing to a stellar encounter \citep[e.g.,][]{Tamayo2016}.


\section{HR 8799 system: flyby of a single Sun} \label{sec-hr8799}

\begin{figure}
\includegraphics[width=\columnwidth]{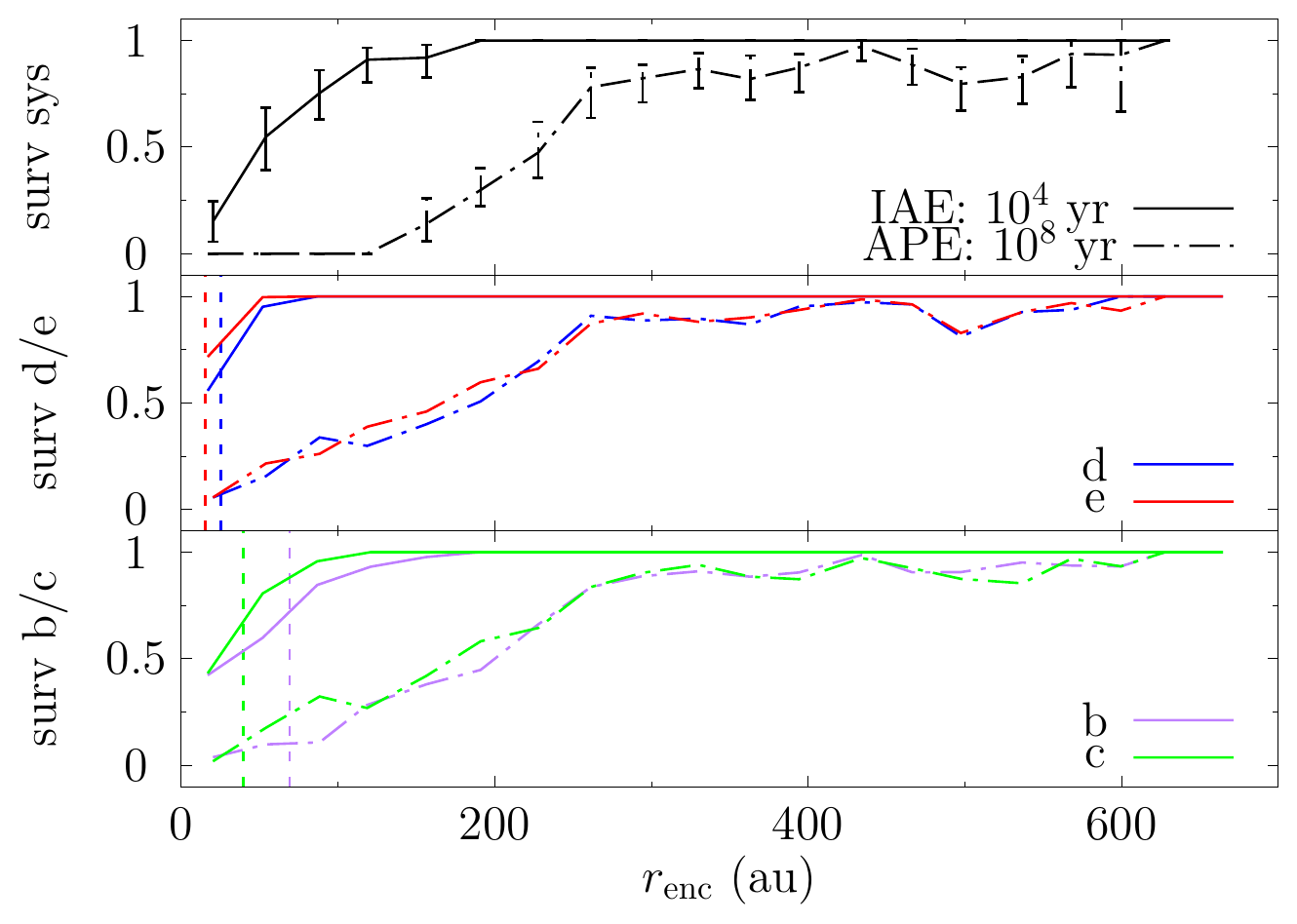}
\caption{Stability of the HR 8799 system at the encounter and post-encounter phases as a function of encounter distance. The encounter distance is normalised such that the outermost planet, b, is at $\sim$70 au. The bottom panels shows the survival fraction of planets b (black) and c (red) at the encounter phase (solid) and the post-encounter phase (dash-dotted); the middle panel show planets d (blue) and e (green). In the top panel, the black line indicates immediate ejection of planets during the encounter, dashed that of ejection plus orbital crossing and dash-dotted for stability at the end of the post-encounter evolution.}
\label{fig-stab-hr8799}
\end{figure}

We first show the survivability of the planets during the encounter and the post-encounter phases in the bottom and middle panels of Figure \ref{fig-stab-hr8799} in different colours. Here, because the semimajor of the outermost planet, b, can vary within the range $\sim 50-80$ au (Section \ref{sec-sys-hr8799}), we have normalised the encounter distance with respect to this value. And when plotting, we assume all systems have b at 69 au, i.e., $r_\mathrm{enc,plot}=r_\mathrm{enc,real}/a_\mathrm{b,real}\times69$ au.

Take the outermost planet, b, for example. During the encounter, it can only be ejected by those $\lesssim 180$ au. During the post-encounter phase, the systems evolve significantly and, this planet is lost much more frequently and at further encounter distances. For example, at $\sim 200$ au, that planet can only survive 50\% of the times and encounters as distant as 600 au, may still cause loss of the planet.

Though during the encounter, the survival fraction of a planet is directly related to its semimajor axis, during the post-encounter phase, such a dependence is removed -- all four planets, irrespective of the position, are equally likely to be disrupted. This agrees with the simulations of the 3J systems (Figure \ref{fig-stab-3j10}) and implies again that in systems of equal-mass planets, the planets' locations do not matter much \citep{Marzari2002}.

The top panel of Figure \ref{fig-stab-hr8799} shows the stability of the system as a whole. As the solid black line shows, the system can only be immediately destabilised by encounters $\lesssim180$ au and the instability fraction reaches 50\% at about 50 au. An encounter may also, while keeping the planets, make their orbits cross. Considering this, now even at 120 au, an encounter still has a chance of 50\% to make the system unstable.

During the post-encounter evolution, the systems are damaged to a much higher extent. None survives an encounter $\lesssim 120$ au and even at 250 au, an encounter still disrupts the system at a chance of 50\%. In rare cases, as discussed for planet b already, an encounter as distant as 600 au (or nine times the stellar-centric distance of the outermost planet) may still be able to induce instability. We do not check the survivability of the resonances but confirm that all stable systems have period ratios close to two.

\subsection{Implication}
The fact that the HR 8799 system can be disrupted by an encounter at hundreds of au means that it probably does not originate from a dense cluster or at least it has not stayed in such a environment in a long time.

The encounters studied in this work are embedded in the open clusters featuring low encounter velocity. But in the field, a star may still, on average, encounter another at a few hundreds of au during its main sequence \citep{Veras2012}. These encounters are weaker because of the much faster encounter velocity \citep{Li2015}. Can the HR 8799 system survive the encounters at 100s of au in the field? This not only has to do the future survivability but also the past. The age of the system is not well constrained and estimate up to 1 giga years was proposed \citep[][this is under debate and perhaps, younger ages of a few tens to a few hundreds of millions of years may be more reasonable \citealt{Moro-Martin2010}]{Moya2010}. If encounters in the field prove detrimental to the system, they can potentially be used to put a upper limit on the system's age.

The formation of wide-orbit massive planets is difficult to explain and gravitational instability (GI) seems to be the only viable mechanism \citep{Dodson-Robinson2009}. In circumstellar disks of $\sim10\%$ of the mass of the central host, gaseous clumps may form and contract to protoplanets at several tens of au within $10^5$ yr \citep{Helled2010,Boss2011}. For this mechanism to work, massive disks must be present. However, in a clustered environment the disks may be subject to truncation. It has been established that truncation due to stellar flybys is weaker than that by photoevaporation \citep{Scally2001,Winter2018}. The latter operates on timescales of $10^5-10^6$ yr \citep{Scally2001,Adams2004,Winter2018}. So the formation of the HR8799 system should not be compromised if GI is the formation mechanism. Moreover, GI tends to produce planets that are too massive to be planets -- it has been suggested \citep{Boss2011} that photoevaporation in a cluster could stop the planet from accreting too much mass. Nonetheless, here our finding implies that even if such multi wide-orbit massive planet systems can form in clusters, they are prone to disruption owing to stellar flybys. Hence, such a system must have been, as a whole, ejected out of the cluster or the cluster itself must have dissolved.



\section{Cross-sections for ejection}
\begin{figure}
\includegraphics[width=\columnwidth]{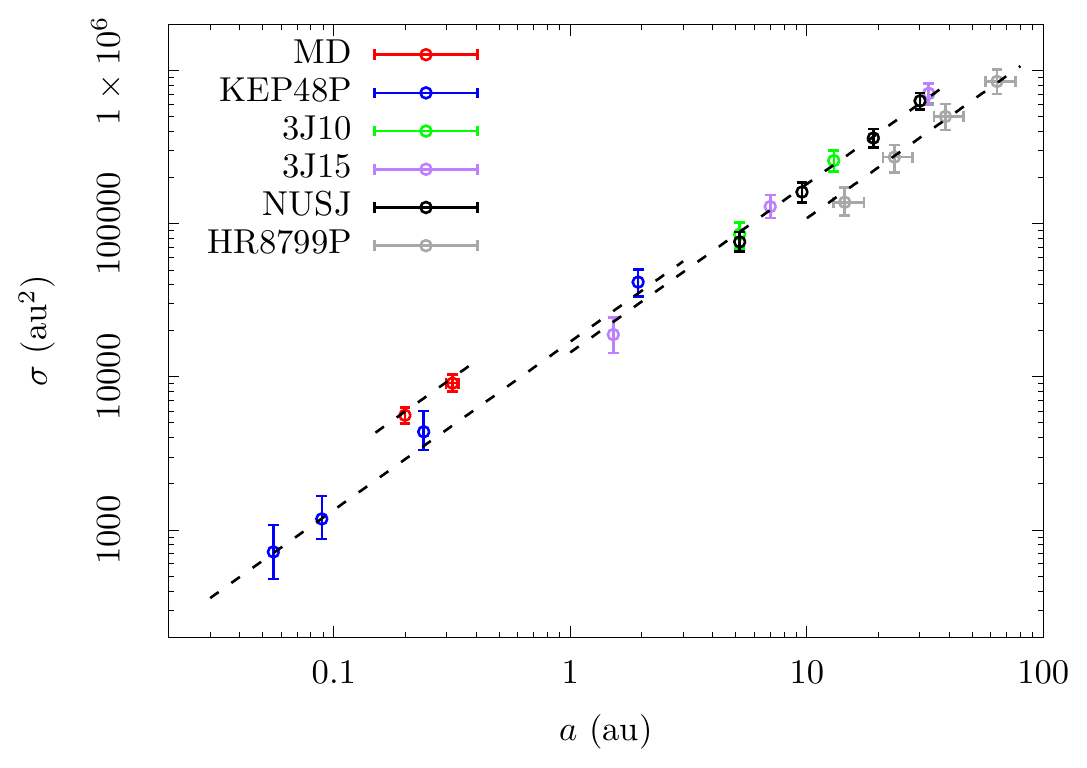}
\caption{Cross section $\sigma$ for the immediate ejection of a planet under the flying-by lone-Sun as a function of the semimajor axis of that planet $a$. Different colours represent planets around different host stars. The dashed lines are a fit of $\log\sigma=c_0+c_1\log m_\mathrm{cen}+c_2\log a$ where $m_\mathrm{cen}$ is the mass of the host star; the fitted parameters are $c_0=4.16$, $c_1=-1.27$ and $c_2=1.10$.}
\label{fig-cross}
\end{figure}

How often the ejection of a planet as an immediate result of a stellar flyby encounter as simulated in this work would occur in an open cluster can be estimated as $\Gamma=n v\sigma$ where $n$ is the number density of the cluster, $v$ the velocity dispersion and $\sigma$ the so-called cross section \citep[e.g.,][]{Hut1983}. The former two depend on the properties of cluster. We have estimated $\sigma$ for all our simulations where the encountering star is the lone Sun following \citetalias{Li2019} and the result is shown in Figure \ref{fig-cross} as a function of the planet's semimajor axis. For the Kepler-48 system where the dynamical timescale due to interplanetary interaction is short, we have used the planetary status when the distance between the Sun and the host star is twice the minimum value and before $10^4$ yr (cf. Figure \ref{fig-stab-kep48-sun}). Taking the NUSJ system for example, the closest planet has a cross section of $8\times10^4$ au$^2$ and it is $6\times10^5$ au$^2$ for the outermost one, in agreement with \citetalias{Li2019}.

Our estimated cross sectional areas \citepalias[also][]{Li2019} are larger than those by \citet{Adams2001,Li2015} by an order of magnitude. We deem that the reason could be: (1) We are using the Sun as the encountering star all the time and the previous authors used stellar masses drawn from a power law distribution. So their intruding stars are on average significantly less massive than ours. (2) We fix the velocity of infinity at 1 km/s while this quantity was varied assuming a Maxwellian distribution in their work. (3) We do not sample the encounter directly in the impact parameter $b$ but create them using the closest approach distance $r_\mathrm{enc}$ assuming that its distribution is flat. This assumption is accurate only for small $r_\mathrm{enc}$ so we could have overestimated $\sigma$ for distant planets. (4) We suspect that the value of $\sigma$ might have not converged in their works. There, the authors were calculating the cross section for ejection caused by a binary (when assuming one component is of zero mass the binary is effective a single star) and they only sampled encounters with impact parameters $b$ smaller than two/ten times the binary separation. Then for pairs much tighter than the planet's semimajor axis $a$, only encounters of $b<a$ were considered. This is probably not sufficient as even at $b\approx 3a$, ejection of the planet is possible \citep{Malmberg2011}. Finally, we note that our result is in general consistent with \citet{Wang2020}.

It seems that in Figure \ref{fig-cross}, for planets around each star, their respective cross sections $\log\sigma$ depend on the semimajor axes $\log a$ linearly, meaning that the actual dependence is a power law; the exponent has been estimated to be unity \citep{Li2015}. These authors also derived that the dependence of on the host star mass followed the same functional form and the exponent was 1/3 \citep[but earlier works also observed 1/2,][]{Adams2006}. Here, for the immediate ejection of a planet at $a$ from a host star of mass $m_\mathrm{cen}$ under the encounter of the Sun with $v_\mathrm{inf}=1$ km/s we fit $\log\sigma=c_0+c_1\log m_\mathrm{cen}+c_2\log a$ and obtain
\begin{equation}
\log\sigma=(4.16\pm0.03)-(1.27\pm0.24)\log m_\mathrm{cen}+(1.10\pm0.03)\log a,
\end{equation}
where $\sigma$ is measured in au$^2$, $m_\mathrm{cen}$ in solar unit and $a$ in au. In Figure \ref{fig-cross}, we have for each stellar mass, over-plotted the fit for the semimajor ranges of the planets under investigation. The agreement is fairly good.

Compared to the literature, our dependence of $\sigma$ on $a$ is only slightly more sensitive than \citet{Li2015}. But that on $m_\mathrm{cen}$ strongly disagrees and our dependence is 3 times steeper. The reason for this is unclear. We comment that the purpose of this work is not to quantify the cross sections but to investigate the evolution of planetary systems after a stellar flyby and the simulations are not optimised for the calculation of $\sigma$.

Finally, we comment that above we have been concerned with the cross section for the immediate ejection only. Later post-encounter evolution would further the degree of instability of the planetary systems, enhancing the cross section. This depends on the specific configuration of the planetary system. In \citetalias{Li2019}, we showed that for the original solar system, $\sigma$ for Uranus could increase by a factor of 3.

\section{Conclusion} \label{sec-con}

We have performed extensive $N$-body simulations to explore the effect of stellar encounters on various types of planetary systems. We have considered both the immediate influence of the encounter and also followed the long-term post-encounter evolution of the systems. Additionally, we have examined the interaction between the captured planets and the original planets. Our main findings are:
\begin{enumerate}
\item Close-in super earth systems are resistant to the direct effect of flyby encounters. To destabilise such a system, an encounter inside a few au is needed. These low mass systems are also stable against self-disruption given lack of instability in the post-encounter evolution. Thus, the only way to damage such system is to do that during the encounter.
\item Mean motion resonances (MMRs) in these systems can only be broken by encounters within several times the semimajor axis of the outer planet, thus still $\lesssim$ a few au. The encounter does not preferentially push the ratio of orbital period toward a specific direction. Thus, the observed imbalance of the distribution of period ratio near first order MMRs is probably not a result of close encounters.
\item Close-in super earths are strongly-coupled with short dynamical timescale. So even if the host star captures a giant planet on a wide orbit from the other star during the encounter, the capture usually cannot destabilise/excite the inner orbits. As a consequence, both the inner original planets themselves and their coplanarity are conserved. But the planets can be tilted as a whole in that while the interplanetary inclinations remain low, their orbits can be highly inclined with respect to the equator of the central star.
\item When these close-in super earths have a outer gas giant neighbour, they are more susceptible to the encounters and now loss during the post-encounter phase is more frequent. Particularly, such systems are more prone to a captured giant planet -- the capture can drive the preexisting gas giant into large-amplitude Kozai--Lidov cycles, hence destroying the original planets.
\item While it is the stellar-centric distances that determine the planets' stability in the encounter phase, their masses play an important role in the post-encounter evolution. In equal-mass planetary systems, the planets have equal chances to be destabilised in the post-encounter evolution while in those with a mass gradient, the most massive one has the best chance to survive.
\item Effects of the encounter may be distinguished from internal scattering. For example, we consider planetary systems, each initially with three equal-mass planets, losing exactly one planet after the post-encounter evolution: those losing one planet during the encounter have colder orbital elements than those losing a planet in the post-encounter phase. Also, in the latter, because interplanetary interactions are the main driver, the resulting orbits resemble those in classical planet scattering simulations without stellar flybys.
\item A stellar flyby can impart angular momentum deficit (AMD) into a planetary system during the encounter. The AMD immediately after the encounter is a key parameter relating to the post-encounter phase evolution. For example, the more the AMD, the more likely the system turns unstable and the earlier. Also, because more AMD can be injected into systems with an outer-heavy mass slope, they are easier to destabilise than those inner-heavy. This may open up the possibility to use AMD immediately after the encounter to predict the behaviour of the system's post-encounter long-term evolution.
\item Systems with multiple massive planets on wide orbits, like that of HR 8799, are especially vulnerable to flyby encounters and one at 250 au still has a chance of 50\% to destabilise such systems. Thus this system cannot have stayed in a clustered environment for a long time.
\end{enumerate}

\section*{Acknowledgements}

The authors are grateful to an anonymous referee for helpful comments. The authors acknowledge financial support from Knut and Alice Wallenberg Foundation through two grants (2014.0017, PI: Melvyn B. Davies and 2012.0150, PI: Anders Johansen). The authors also thanks the Royal Physiographic Society of Lund. Computations were carried out at the center for scientific and technical computing at Lund University (LUNARC) through the Swedish National Infrastructure for Computing (SNIC) via project 2019/3-398. Figure \ref{fig-system_state} is made using \url{http://sankeymatic.com/}.


\bsp	
\label{lastpage}
\end{document}